\documentclass{aa}

\usepackage{graphicx}
\usepackage{threeparttable}
\usepackage{txfonts}
\usepackage{color}
\usepackage{lscape}
\usepackage{afterpage}
\usepackage{ulem}
\usepackage{comment}
\begin{document}

   \title{Constraints on PDS~70~b and c from the dust continuum emission of the circumplanetary discs considering in situ dust evolution}

   \titlerunning{Constraints on PDS~70~b and c from the dust emission of the CPDs} 
    

   \author{Y. Shibaike\inst{1,2}
          \and
          C. Mordasini\inst{1}
          }

   \institute{Space Research and Planetary Sciences, Physics Institute, University of Bern, CH-3012 Bern, Switzerland
   \and
   National Astronomical Observatory of Japan, 2-21-1 Osawa, Mitaka, Tokyo 181-8588, Japan
   \\
              \email{yuhito.shibaike@nao.ac.jp}
             }

   \date{Received MM. DD, 2024; accepted MM DD, 2024}

 
  \abstract
   {The young T Tauri star PDS~70 has two gas accreting planets sharing one large gap in a pre-transitional disc. Dust continuum emission from PDS~70~c has been detected by Atacama Large Millimeter/submillimeter Array (ALMA) Band 7, considered as the evidence of a circumplanetary disc. However, there has been no detection of the dust emission from the CPD of PDS~70~b.}
   {We constrain the planet mass and the gas accretion rate of the planets by introducing a model of dust evolution in the CPDs and reproducing the detection and non-detection of the dust emission.}
   {We first develop a 1D steady gas disc model of the CPDs reflecting the planet properties. We then calculate the radial distribution of the dust profiles considering the dust evolution in the gas disc and calculate the total flux density of dust thermal emission from the CPDs.}
   {We find positive correlations between the flux density of dust emission and three planet properties, the planet mass, gas accretion rate, and their product called `MMdot'. We then find that the MMdot of PDS~70~c is $\geq4\times10^{-7}~M_{\rm J}^{2}~{\rm yr}^{-1}$, corresponding to the planet mass of $\geq5~M_{\rm J}$ and the gas accretion rate of $\geq2\times10^{-8}~M_{\rm J}~{\rm yr}^{-1}$. This is the first case to succeed in obtaining constraints on planet properties from the flux density of dust continuum emission from a CPD. We also find some loose constraints on the properties of PDS~70~b from the non-detection of its dust emission.}
   {We propose possible scenarios for PDS~70~b and c explaining the non-detection respectively detection of the dust emission from their CPDs. The first explanation is that planet c has larger planet mass, larger gas accretion rate, or both than planet b. The other possibility is that the CPD of planet c has a larger amount of dust supply, weaker turbulence, or both than that of planet b. If the dust supply to planet c is larger than b due to its closeness to the outer dust ring, it is also quantitatively consistent with that planet c has weaker H$\alpha$ line emission than planet b considering the dust extinction effect.
   }


   \maketitle
   
%
\nolinenumbers
\section{Introduction} \label{introduction}
Forming planets with enough mass embedded in protoplanetary discs (PPDs) accrete gas from the discs and form small gas discs called circumplanetary discs (CPDs) around them. There have been a lot of (magneto-) hydrodynamical ((M)HD) simulations of gas accreting planets to reveal the gas accreting process, one of the most fundamental processes of giant planet formation \citep[e.g.][]{lub99,tan12,gre13,sch20}. In addition, CPDs are the birthplaces of large satellites around gas planets; therefore the discs have been investigated in the context of the satellite formation \citep[e.g.][]{can06,shi19}. However, there were no detection of gas accreting planets nor CPDs in extrasolar systems until just recently, meaning that their research had been restricted to theoretical approaches such as numerical simulations.

Recently, several gas accreting planets and a CPD have been discovered, making the subject highly interested. There have been two gas accreting planets reported around a young T Tauri star PDS~70 (spectral type K7; $M_{\rm star}=0.76~M_{\odot}$; $5.4~{\rm Myr}$ old), where the system is located at $d=113.43~{\rm pc}$ in the Upper Centaurus Lupus association \citep{gaia18,mul18}. PDS~70~b and c are located at a semi-major axis of $a_{\rm pl}=20.6$ and $34.5~{\rm au}$, respectively, and share a large gap in a pre-transitional disc with an inclination of $i=51.7^{\circ}$ \citep{mul18,kep19,haf19}. The two planets have been observed in many ways such as multiple infrared (IR) wavelengths and H$\alpha$ emission \citep[e.g.][]{kep18,mul18,aoy19,haf19,wan21}. These observations can constrain two important properties of the forming planets: the planet mass ($M_{\rm pl}$), the gas accretion rate ($\dot{M}_{\rm g,pl}$), or both. The flux of such emission from the planet b is higher than that of c in most of the previous observations, suggesting that the planet mass and the gas accretion rate of the planet b are higher than those of c (see also Tab. \ref{tab:estimates} and Section \ref{constraints} for more detailed explanations of the previous observations).

On the other hand, there has been only one detection of a CPD: the dust continuum emission  from the CPD around PDS~70~c with the Atacama Large Millimeter/submillimeter Array (ALMA) in Band 7 ($\lambda=855~\mu{\rm m}$) \citep{ise19,ben21,cas22}\footnote{\citet{chr19b} shows that the best fit to the SED (IR) of PDS~70~b is obtained by a model considering a CPD around the planet. There is also a candidate CPD in the system of AS209 discovered by the distortion of the gas velocity field \citep{bae22}. A few marginal CPD candidates are also discovered by the dust continuum of $\lambda=1.25~{\rm mm}$ in the Disk Substructures at High Angular Resolution Project (DSHARP) survey \citep{and21}.}. The dust emission has not been detected from the CPD of PDS~70~b but only from the predicted location of $L_{5}$ of the planet \citep{ben21,bal23}. This fact that only the planet c has the detection of the dust continuum looks inconsistent with that the IR and H$\alpha$ luminosity of the planet b is higher than that of c, which has been one of the issues not solved yet. \citet{ben21} detected $I_{\rm d}=86\pm16~\mu{\rm Jy}~{\rm beam}^{-1}$ dust continuum emission (peak intensity) from PDS~70~c with the noise level of $1\sigma=15.7~\mu{\rm Jy}$. The radius of CPD is about $r_{\rm out}=1/3~R_{\rm H}$, which is about $1~{\rm au}$ in the case of PDS~70~c, meaning that it is difficult to resolve the CPD by ALMA and any other current telescopes. The Hill radius is defined as $R_{\rm H}\equiv\{M_{\rm pl}/(3M_{\rm star})\}^{1/3}a_{\rm pl}$, where $M_{\rm pl}$, $M_{\rm star}$, and $a_{\rm pl}$ are the planet mass, the stellar mass, and the orbital distance of the planet, respectively. Therefore, only information we can obtain from the dust continuum observation is its total flux density from the CPD ($F_{\rm emit}$). We note that \citet{cas22} revisits the ALMA data and finds that the flux density of dust emission from PDS~70~c could be variable by at least $42\pm13\%$ over a few years time-span.

\citet{ben21} and the other previous works \citep[e.g.][]{bae19} estimate the (maximum) dust size ($R_{\rm d}$) and the dust mass ($M_{\rm d}$) from the observed value of $F_{\rm emit}$. If the size of all dust in the CPD is $R_{\rm d}=1~{\rm mm}$ and $1~\mu{\rm m}$, the dust mass is estimated as $M_{\rm d}\sim0.007M_{\oplus}$ and $\sim0.031M_{\oplus}$, respectively \citep{ben21}. However, the evolution of dust is not considered in the previous works, meaning that the two parameters are free parameters, resulting in that the degeneracy of the two parameters is impossible to be resolved. Here, we introduce a dust evolution model developed in the context of the satellite formation in CPDs and replace the two parameters, $R_{\rm d}$ and $M_{\rm d}$, to one single parameter, $\dot{M}_{\rm d,tot}$, the total mass flux of dust inflowing to the CPDs \citep{shi17,shi23}. As a result, $R_{\rm d}$ and $M_{\rm d}$ can be calculated from $\dot{M}_{\rm d,tot}$ and the conditions of the gas discs. The conditions of the viscous accretion CPDs (i.e. the radial profiles of the gas surface density $\Sigma_{\rm g}$ and midplane temperature $T_{\rm mid}$) are determined by the three dominant parameters: the planet mass ($M_{\rm pl}$), the mass flux of gas inflowing to the CPDs ($\dot{M}_{\rm g,tot}(\approx\dot{M}_{\rm g,pl})$) and the strength of turbulence in the CPDs ($\alpha$). Moreover, the realistic value of the parameter $\dot{M}_{\rm d,tot}$ is not so wide, because the dust-to-gas mass ratio in the gas inflow to CPDs, $x\equiv\dot{M}_{\rm d,tot}/\dot{M}_{\rm g,tot}$, should be lower than the stellar composition $0.01$ by considering the effect of dust filtering at the edge of the gap the two planets sharing. Therefore, there is a possibility to constrain the planet properties, $M_{\rm pl}$ and $\dot{M}_{\rm g,pl}$, from the observed dust thermal emission value, $F_{\rm emit}=86\pm16~\mu{\rm Jy}$.

In Section \ref{methods}, we explain the models and parameters used in this work. Sections \ref{disc}, \ref{evolution}, and \ref{emission} describe the gas disc (CPD), dust evolution, and dust continuum emission models, respectively. We summarise the parameter settings in Section \ref{parameters}. In Section \ref{results}, we first show the examples of the radial profiles of the gas and dust in CPDs calculated by our model explained in Section \ref{distribution}. We then investigate the dependence of the flux density of dust emission from CPDs on the planet mass and the gas accretion rate (Section \ref{planet-properties}). We also investigate the effects of the other properties of the planets and CPDs in Section \ref{properties}. We then obtain the constraints on the properties of PDS~70~b and c by using the revealed dependence and compare our estimates with those of the previous works (Section \ref{constraints}). In Section \ref{scenario}, we propose possible scenarios for PDS~70~b and c consistent with the constraints obtained in the previous section. In Section \ref{conclution}, we conclude our research. We also explain some detailed parts of the models in Appendix.

\section{Methods} \label{methods}
\subsection{Gas disc model} \label{disc}
\begin{figure}[tbp]
\centering
\includegraphics[width=0.9\linewidth]{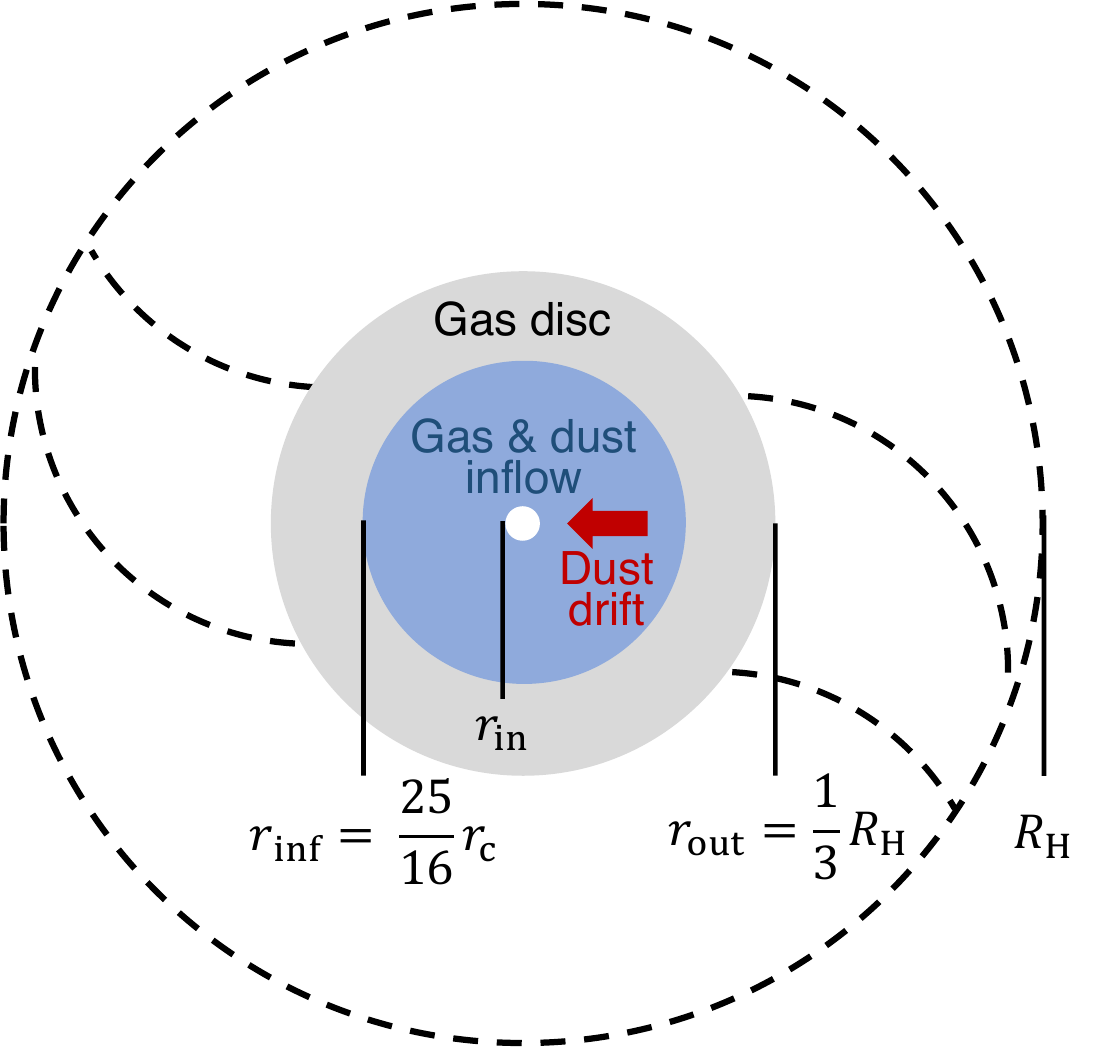}
\caption{Schematic picture of the disc model. Gas and dust flow onto the region $r\leq r_{\rm inf}$ of the CPD with a uniform mass flux (blue region). The gas disc expands outwards by the diffusion and is truncated at $r=r_{\rm out}$. The disc is also truncated by the magnetospheric cavity of the planet at $r=r_{\rm in}$. Dust drifts towards the planet and only exists in the blue region.}
\label{fig:models}
\end{figure}

In the previous works of the observations of CPDs and forming planets, simple 1D disc models have been used for CPDs \citep{zhu15a,eis15}. In such models, the gas surface density has a power-low radial distribution assuming viscous accretion. Here, we model a detailed steady 1D gas disc with gas inflow based on the model proposed in \citet{can02} called as the `gas-starved' disc model. Figure \ref{fig:models} is the schematic picture of our gas (and dust) disc model. Unlike the assumptions in the previous work, we use an expression about the specific angular momentum of the inflowing gas derived from the results of multiple previous hydrodynamical simulations, which determines the position of the outer edge of the gas inflow region \citep{war10}. We also introduce the inner edge of the disc due to the magnetic field of the planet. Plus, we calculate the disc temperature more detailed than the previous work. We iteratively calculate the gas and temperature profiles simultaneously. We explain the model in detail in the following sections.

\subsubsection{Inflow and surface density of the disc} \label{gas}
First, we assume that the outer edge of the disc as $r_{\rm out}=1/3~R_{\rm H}$. Most previous hydrodynamical simulations show that the gas structure is azimuthally symmetric inside the outer edge, and the gas (surface) density of $r<r_{\rm out}$ is much higher than that of $r>r_{\rm out}$ \citep[e.g.][]{tan12,sch20}. Therefore, we only calculate the gas structure inside $r_{\rm out}$ by a 1D (radial direction) disc model.

We consider a steady gas disc with constant supply of gas. We assume that the gas flows into the region of $r_{\rm in}\leq r\leq r_{\rm inf}$, where $r$ is the distance from the planet, $r_{\rm in}$ is the position of the inner edge of the disc (see Section \ref{inneredge}), and $r_{\rm inf}$ is the outer boundary of the gas inflow region. We assume that the gas flows onto the disc with uniform mass flux per area, $F_{\rm g}=\dot{M}_{\rm g,tot}/(\pi r_{\rm inf}^{2})$, where $\dot{M}_{\rm g,tot}$ is the total mass rate of the gas inflow onto the CPD\footnote{The $r$ dependence of the mass flux is still controversial. For example, a hydrodynamical simulation by \citet{tan12} finds about $\propto r^{-1}$.}. In that case, $r_{\rm inf}=25/16~r_{\rm c}$, where $r_{\rm c}\equiv j_{\rm c}^{2}/(GM_{\rm pl})$ is the centrifugal radius of the inflowing gas with the average specific angular momentum, $j_{\rm c}$ \citep{can02,war10}. The letter $G$ is the gravitational constant. We then define the average gas specific angular momentum as $j_{\rm c}\equiv l\Omega_{\rm K,pl}R_{\rm H}^{2}$, where $l$ is the angular momentum bias, and $\Omega_{\rm K,pl}=\sqrt{GM_{\rm pl}/a_{\rm pl}^{3}}$ is the Keplerian frequency of the planet. There is a correlation between the centrifugal and Hill radii that $r_{\rm c}=l^{2}/3~R_{\rm H}$. When $l=1$, the centrifugal radius reaches the outer edge of the disc (i.e. $r_{\rm c}=r_{\rm out}$). On the other hand, the value $l=1/4$ corresponds to the specific angular momentum of undeflected 2D Keplerian flow (i.e. neglecting the gravity of the planet) across the region of $r=R_{\rm H}$ (i.e, accretion boundary) \citep{lis91,war10}\footnote{We assume that the Bondi radius, $R_{\rm B}$, is larger than the Hill radius. Otherwise, the gas across the region between the two radii does not accrete onto the CPD, and $l=1/4$ should be corrected to $l=1/4(R_{\rm B}/R_{\rm H})^{2}$ \citep{war10}.}. In our model, we use an approximation to calculate $l$ derived by previous hydrodynamical simulations \citep{war10}:
\begin{equation}
l=0.12\left(\dfrac{R_{\rm B}}{R_{\rm H}}\right)^{1/2}+0.13,
\label{l}
\end{equation}
where $R_{\rm B}=GM_{\rm pl}/c_{\rm s,PPD}^{2}$ is the Bondi radius. Here, we use $c_{\rm s,PPD}=\sqrt{k_{\rm B}T_{\rm PPD}/\mu m_{\rm H}}$ as the isothermal sound speed around the planet, where $k_{\rm B}$, $T_{\rm PPD}$, and $m_{\rm H}$ are the Boltzmann constant, the temperature of the protoplanetary disc around the planet, and the mass of an hydrogen atom, respectively. We assume the mean molecular weight of the gas as $\mu=((1-Y)/2.006+Y/4.008)^{-1}=2.32$ with the mass fraction of helium of $Y=0.27$. We note that, however, these previous numerical simulations assume the cases of Jupiter (or similar planets), and the ratio $R_{\rm B}/R_{\rm H}$ of PDS~70~b and c are much larger than what they assume.

The steady state gas surface density profile of a viscous accretion disc is analytically solved \citep{can02},
\begin{equation}
\Sigma_{\rm g,CW}=\dfrac{4\dot{M}_{\rm g,tot}}{15\pi\nu}
\begin{cases}
\dfrac{5}{4}-\sqrt{\dfrac{r_{\rm inf}}{r_{\rm out}}}-\dfrac{1}{4}\left(\dfrac{r}{r_{\rm inf}}\right)^{2}, & r<r_{\rm inf}, \\
\sqrt{\dfrac{r_{\rm inf}}{r}}-\sqrt{\dfrac{r_{\rm inf}}{r_{\rm out}}}, & r\geq r_{\rm inf},
\end{cases}
\label{SigmagCW}
\end{equation}
where $r_{\rm in}\ll r_{\rm inf}$ and $\nu$ is the disc viscosity. The total mass flux onto the planet (i.e, the gas accretion rate onto the planet) through the disc is,
\begin{equation}
\dot{M}_{\rm g,pl}=\dot{M}_{\rm g,tot}\left(1-\dfrac{4}{5}\sqrt{\dfrac{r_{\rm inf}}{r_{\rm out}}}\right).
\label{Mdodtgpl}
\end{equation}
When the innermost region of the disc is truncated at $r_{\rm in}$, the gas surface density is corrected to
\begin{equation}
\Sigma_{\rm g}=\Sigma_{\rm g,CW}\left(1-\sqrt{\dfrac{r_{\rm in}}{r}}\right)\left(1-\sqrt{\dfrac{r_{\rm in}}{r_{\rm out}}}\right)^{-1}.
\label{Sigmag}
\end{equation}
We do not consider any other sub-structures of CPDs such as pressure bumps formed by potential satellites, which may increase the flux density of dust emission with halting the dust drift \citep{bae19}. However, there are numerous of possibilities, it is beyond the scope of this paper to consider them.

The gas surface density depends on the midplane temperature of the disc through the disc viscosity, $\nu\equiv\alpha c_{\rm s}H_{\rm g}$, where $\alpha$, $c_{\rm s}$, and $H_{\rm g}$ are the strength of turbulence (assumed uniform), the isothermal sound speed, and the gas scale height, respectively \citep{sha73}. The isothermal sound speed depends on the midplane temperature, $T_{\rm mid}$, with $c_{\rm s}=\sqrt{k_{\rm B}T_{\rm mid}/(\mu m_{\rm H})}$. We calculate the midplane temperature in Section \ref{temperature}. The gas scale height is $H_{\rm g}=c_{\rm s}/\Omega_{\rm K}$, where $\Omega_{\rm K}=\sqrt{GM_{\rm pl}/r^{3}}$ is the Keplerian frequency.

\subsubsection{Inner edge of the disc} \label{inneredge}
The position of the disc inner edge, $r_{\rm in}$, is assumed to depend on the strength of the magnetic field of the planet via a magnetospheric cavity. There is a correlation between the strength of the magnetic field of planets (and stars) and the energy flux available for generating the field \citep{chr09}. The strength of the magnetic field at the surface of the planet (on the midplane) is,
\begin{equation}
B_{\rm s}=\dfrac{1}{f_{\rm surf}}\sqrt{2\mu_{0}cf_{\rm ohm}\langle\rho_{\rm dyn}\rangle^{1/3}\left(\dfrac{F}{q_{0}}\right)^{2/3}}
\label{Bplsurf}
\end{equation}
where $\mu_{0}=4\pi$ is permeability, $\langle\rho_{\rm dyn}\rangle$ is the mean density of the dynamo region, and $q_{0}=L_{\rm dyn}/(4\pi R_{\rm dyn}^{2})$ is the bolometric flux at the outer boundary of the dynamo region. Here, we assume that the top of the dynamo region is corresponding to the surface of the planet (i.e. $R_{\rm dyn}=R_{\rm pl}$); therefore $L_{\rm dyn}=L_{\rm pl}$, where $L_{\rm pl}$ is the luminosity of the planet, and $\langle\rho_{\rm dyn}\rangle=M_{\rm pl}/(4\pi/3 R_{\rm pl}^{3})$. The planet radius, $R_{\rm pl}$, is an input parameter. We set the factor from the mean internal magnetic field strength of the dynamo region $\langle B_{\rm dyn}\rangle$ to the mean surface field $B_{\rm s}$ as $f_{\rm surf}=\langle B_{\rm dyn}\rangle/B_{\rm s}=3.5$, the constant of proportionality as $c=0.63$, the ratio of ohmic dissipation to total dissipation as $f_{\rm ohm}=1$, and the efficiency factor considering the averaging of the radially varying properties as $F=1$ (see \citet{chr09} for the details).

The disc inner edge, in other words, the truncation radius by the magnetospheric cavity of the planet is
\begin{equation}
r_{\rm in}=\left(\dfrac{\mu'\mathcal{M}^{2}}{2\Omega(r_{\rm in})\dot{M}_{\rm g,pl}}\right)^{1/5}=\left(\dfrac{\mathcal{M}^{4}}{4GM_{\rm pl}\dot{M}_{\rm g,pl}^{2}}\right)^{1/7}
\label{rin}
\end{equation}
where $\mathcal{M}=B_{\rm s}R_{\rm pl}^{3}$, and the magnetic field is assumed to be a pole-aligned dipole \citep{dan10,tak22}. This expression is derived from the condition for the steady angular momentum transfer at the disc inner edge, $\dot{M}_{\rm g,pl}\Omega(r_{\rm in})\approx B_{\phi}B_{\rm z}r_{\rm in}$, where MHD simulations by \citet{tak22} show that $\Omega(r_{\rm in})=\Omega_{\rm K}(r_{\rm in})$ and $\mu'=|B_{\phi}/B_{\rm z}|=1$ in the case of a T Tauri star. Note that this expression is the same with the classical expression derived from the balance between the magnetic and ram pressure with spherical accretion except for small difference by a factor of a few \citep{gho79}.

\subsubsection{Disc temperature} \label{temperature}
We calculate the disc temperature as follows instead of the simplified way used in \citet{can02}. The midplane temperature $T_{\rm mid}$ can be calculated from the energy balance between each heat source and sink \citep{nak94,hue05,ems21a},
\begin{equation}
\sigma_{\rm SB}T_{\rm mid}^{4}=\dfrac{1}{2}\left\{\left(\dfrac{3}{8}\tau_{\rm R}+\dfrac{1}{2\tau_{\rm P}}\right)\dot{E}_{\rm v}+\left(1+\dfrac{1}{2\tau_{\rm P}}\right)\dot{E}_{\rm s}\right\}+\sigma_{\rm SB}T_{\rm irr,tot}^{4},
\label{Tmid}
\end{equation}
where $\sigma_{\rm SB}$ is the Stefan-Boltzmann constant, $\tau_{\rm R}$ and $\tau_{\rm P}$ are the Rosseland and Planck mean optical depths, $\dot{E}_{\rm v}$ and $\dot{E}_{\rm s}$ are the viscous dissipation rate and the shock-heating rate per unit surface area, and $T_{\rm irr,tot}$ is the total effective temperature heated by irradiation from multiple heat sources.

The Rosseland mean optical depth is $\tau_{\rm R}=\kappa_{\rm disc}\Sigma_{\rm g}$, where we calculate the opacity $\kappa_{\rm disc}(\rho_{\rm g,mid}, T_{\rm mid})$ as the maximum of the fixed dust opacity computed by \citet{bel94}, considering the dust-to-gas surface density ratio ($Z_{\rm \Sigma,est}$) with an approximate expression consistent with the ratio calculated in Section \ref{evolution} (see Appendix \ref{opacity} for the detail), and the gas opacity calculated by \citet{fre14}. The gas density on the midplane is $\rho_{\rm g,mid}=\Sigma_{\rm g}/(\sqrt{2\pi}H_{\rm g})$. We set the Planck mean optical depth as $\tau_{\rm P}=2.4\tau_{\rm R}$ \citep{nak94}.

The viscous dissipation rate is
\begin{equation}
\dot{E}_{\rm v}=\Sigma_{\rm g}\nu\left(r\dfrac{\partial\Omega_{\rm K}}{\partial r}\right)^{2}=\dfrac{9}{4}\Sigma_{\rm g}\nu\Omega_{\rm K}.
\label{dotEv}
\end{equation}
The shock-heating rate is
\begin{equation}
\dot{E}_{\rm s}=
\begin{cases}
\dfrac{GM_{\rm pl}\dot{M}_{\rm g,tot}}{\pi r_{\rm inf}^{2}}\left(\dfrac{1}{r}-\dfrac{1}{r_{\rm out}}\right), & r\leq r_{\rm inf}, \\
0, & r_{\rm inf}<r,
\end{cases}
\label{dotEs}
\end{equation}
where we assume it is equal to the rate of gravitational energy released when the gas falls onto the disc with free-fall from the position (altitude) where the distance to the planet is $r_{\rm out}$.

The total effective temperature resulting from irradiation is calculated by
\begin{equation}
T_{\rm irr,tot}^{4}=T_{\rm irr,surf}^{4}+T_{\rm irr,mid}^{4}+T_{\rm irr,PPD}^{4},
\label{Tirrtot}
\end{equation}
where $T_{\rm irr,surf}$, $T_{\rm irr,mid}$, and $T_{\rm irr,PPD}$ are the effective temperature heated by the irradiation from the planet (the surfaces of the disc are heated), heated directly by the irradiation of the planet through the midplane, and heated by the surrounding PPD, respectively. The first term of the right side of Eq. (\ref{Tirrtot}) is,
\begin{equation}
T_{\rm irr,surf}^{4}=T_{\rm pl}^{4}\left\{\dfrac{2}{3\pi}\left(\dfrac{R_{\rm pl}}{r}\right)^{3}+\dfrac{1}{2}\left(\dfrac{R_{\rm pl}}{r}\right)^{2}\dfrac{H_{\rm g}}{r}\left(\dfrac{\partial \ln{H_{\rm g}}}{\partial \ln{r}}-1\right)\right\}.
\label{Tirrsurf}
\end{equation}
The first and second terms in the bracket represent the irradiation onto flat and flaring discs, respectively. We do not directly calculate $\partial \ln{H_{\rm g}}/\partial \ln{r}$ but give a fixed value of $9/7$ \citep{chi97}. The planet temperature is
\begin{equation}
T_{\rm pl}=\left(\dfrac{L_{\rm pl,tot}}{4\pi\sigma_{\rm SB}R_{\rm pl}}\right)^{1/4},
\label{Tpl}
\end{equation}
where $L_{\rm pl,tot}=L_{\rm pl}+L_{\rm shock}$ is the total luminosity of the planet. The intrinsic planet luminosity, $L_{\rm pl}$, is an input parameter and is corresponding to the effective temperature of the planet, $T_{\rm eff}$, with $L_{\rm pl}\equiv4\pi\sigma_{\rm SB}R_{\rm pl}^{2}T_{\rm eff}^{4}$. The luminosity by the shock created by the gas accretion onto the planet is
\begin{equation}
L_{\rm shock}=\eta_{\rm eff}\dfrac{GM_{\rm pl}\dot{M}_{\rm g,pl}}{R_{\rm pl}}\left(1-\dfrac{R_{\rm pl}}{r_{\rm in}}\right),
\label{Lshock}
\end{equation}
where we fix the global radiation efficiency of the gas accretion shock as $\eta_{\rm eff}=0.95$ \citep{mar19}. The second term of Eq. (\ref{Tirrtot}) is
\begin{equation}
T_{\rm irr,mid}^{4}=\dfrac{L_{\rm pl,tot}}{16\pi r^{2}\sigma_{\rm SB}}\exp{(-\tau_{\rm mid})},
\label{Tirrmid}
\end{equation}
where the horizontal optical depth thorough the midplane is
\begin{equation}
\tau_{\rm mid}=\int^{r}_{r_{\rm in}}\rho_{\rm g,mid}\kappa_{\rm disc}~dr.
\label{taumid}
\end{equation}
We note that this term is not important in the situations considered in this work; it is important only in the very final stage of the disc. The third term of Eq. (\ref{Tirrtot}), $T_{\rm irr,PPD}$ is equal to the temperature in the surrounding PPD, $T_{\rm PPD}$, which is an input parameter of this model. In this work, we use the value $51~{\rm K}$ (PDS~70~b) and $32~{\rm K}$ (PDS~70~c) obtained by substituting their orbital radii for the temperature model provided in \citet{law24}, which fits to observations in a set of CO isotopologue lines.

\subsection{Evolution of dust particles} \label{evolution}
We calculate the growth and drift of dust particles in the fixed 1D gas discs model of Section \ref{disc}. We calculate the radial distribution of the surface density of the dust particles, $\Sigma_{\rm d}$, and their peak mass, $m_{\rm d}$, by solving the following equations, Eqs. (\ref{Mdotdust}) and (\ref{growth}), simultaneously. Here, we implicitly assume that the evolution timescale of the particles is much shorter than that of the gas in order to treat the dust (and gas) distribution as steady.

We assume the size of supplied dust as $R_{\rm d,0}=1~\mu{\rm m}$, because only such a small dust can penetrate inside the gap of PDS~70 system by being coupled with gas (see Figure 4 of \citet{bae19}). If the dust is well coupled even with the gas inflow onto CPDs, the dust-to-gas mass flux ratio of the inflow should be assumed uniform in the whole inflow region, $r\leq r_{\rm inf}$. Also, we only consider the dust in $r_{\rm in}\leq r\leq r_{\rm inf}$, because we assume that the supplied dust only moves inwards in CPDs. Then, from the conservation of mass, the dust mass accretion rate inside the CPDs is
\begin{equation}
\dot{M}_{\rm d}=\dot{M}_{\rm d,tot}\left\{1-\left(\dfrac{r}{r_{\rm inf}}\right)^{2}\right\}=-2\pi rv_{r}\Sigma_{\rm d},
\label{Mdotdust}
\end{equation}
where $\dot{M}_{\rm d,tot}$ is the total dust mass flux flowing onto the CPD from the parental PPD. Here, we define the dust-to-gas mass ratio in the gas inflow as $x\equiv\dot{M}_{\rm d,tot}/\dot{M}_{\rm g,tot}$, which is one of the most important parameters in this work. We assume that, inside the snowline (see Eqs (\ref{PevH2O}) and (\ref{PH2O})), $\dot{M}_{\rm d}$ becomes half of that outside the snowline because the water ice evaporates from the dust particles. We note that if the dust growth timescale is not quick enough, the size frequency distribution of dust may have two peaks: a peak composed of the particles drifting as pebbles and that of the particles just supplied to the CPDs. However, we show that such small grains will not play an important role in the total millimeter flux density of dust emission by changing the slope of the dust size frequency distribution and making sure it does not affect the results a lot in Section \ref{parameters}.

The collisional growth of the drifting particles in CPDs is \citep{sat16},
\begin{equation}
v_{\rm r}\dfrac{d m_{\rm d}}{d r}=\epsilon_{\rm grow}\dfrac{2\sqrt{\pi}R_{\rm d}^{2}\Delta v_{\rm dd}}{H_{\rm d}}\Sigma_{\rm d},
\label{growth}
\end{equation}
where $\epsilon_{\rm grow}$, $\Delta v_{\rm dd}$, and $H_{\rm d}$ are the sticking efficiency for a single collision, collision velocity, and vertical dust scale height, respectively. The mass of a single dust particle is $m_{\rm d}=(4\pi/3) R_{\rm d}^{3}\rho_{\rm int}$, where $\rho_{\rm int}=1.4$ and $3.0~{\rm g~cm^{-3}}$ are the internal density of the icy and rocky particles, respectively. In this work, we assume that the particles are compact. Even if the particles are fluffy, the radial distribution of the particles will not change so much \citep{shi17,shi23}, but the dust emission could change \citep{kat14}, and the investigation of its effect is a future work.

The Stokes number of particles in the Epstein, Stokes, and Newton regimes can be expressed by a single equation \citep{ron17},
\begin{equation}
{\rm St}=\left\{\dfrac{\rho_{\rm g,mid}v_{\rm th}}{\rho_{\rm int}R_{\rm d}}\min\left(1,\dfrac{3}{8}\dfrac{\Delta v_{\rm dg}}{v_{\rm th}}C_{\rm D}\right)\right\}^{-1}\Omega_{\rm K},
\label{St}
\end{equation}
where $v_{\rm th}=\sqrt{8/\pi}c_{\rm s}$ is the thermal gas velocity, $\Delta v_{\rm dg}$ is the relative velocity between the dust particles and gas, and $C_{\rm D}$ is a dimensionless coefficient that depends on the particle Reynolds number, ${\rm Re_{p}}$. The particle Reynolds number is
\begin{equation}
{\rm Re_{p}}=\dfrac{4R_{\rm d}\Delta v_{\rm dg}}{v_{\rm th}\lambda_{\rm mfp}},
\label{Rep}
\end{equation}
where $\lambda_{\rm mfp}=m_{\rm g}/(\sigma_{\rm mol}\rho_{\rm g,mid})$ is the mean free path of the gas molecules with their collisional cross section being $\sigma_{\rm mol}=2\times10^{-15}{\rm cm}^{2}$. The critical particle Reynolds number is ${\rm Re_{p}}=24/C_{\rm D}$, where we calculate $C_{\rm D}$ as \citep{per11},
\begin{equation}
C_{\rm D}=\dfrac{24}{\rm Re_{p}}\left(1+0.27{\rm Re_{p}}\right)^{0.43}+0.47\left[1-\exp\left(-0.04{\rm Re_{p}}^{0.38}\right)\right].
\label{CD}
\end{equation}

The dust diffusion determines the vertical distribution except for the case that the diffusion is weak, and the Kelvin-Helmholtz (KH) instability plays a role \citep[e.g.][]{chi10}. The scale height induced by the vertical diffusion is \citep{you07},
\begin{equation}
H_{\rm d,diff}=H_{\rm g}\left(1+\dfrac{\rm St}{\alpha_{\rm diff}}\dfrac{1+2{\rm St}}{1+{\rm St}}\right)^{-1/2}.
\label{Hddiff}
\end{equation}
The scale height induced by the KH instability is,
\begin{equation}
\begin{split}
H_{\rm d,KH}&={\rm Ri}^{1/2}\dfrac{Z_{\rho}^{1/2}}{(1+Z_{\rho})^{3/2}}\eta r \\
&=\{{\rm Ri}Z_{\Sigma}H_{\rm g}(\eta r)^{2}\}^{1/3} - Z_{\Sigma}H_{\rm g}
\label{HdKH}
\end{split}
\end{equation}
where ${\rm Ri}=0.5$ is the Richardson number for the particles, and $Z_{\rho}=\rho_{\rm d,mid}/\rho_{\rm g,mid}$ is the dust-to-gas midplane density ratio \citep{hyo21a}. The KH instability gives the minimum dust scale height when the particles are small, which is assumed as ${\rm St}<1$, but the instability does not grow when the particles are large \citep{mic06}. Then, the dust scale height is calculated as,
\begin{equation}
H_{\rm d}=
\begin{cases}
\max\{H_{\rm d,diff}, H_{\rm d,KH}\}, & {\rm St}<1, \\
H_{\rm d,diff}, & 1\leq{\rm St}.
\end{cases}
\label{Hd}
\end{equation}
The midplane dust density is $\rho_{\rm d,mid}=\Sigma_{\rm d}/(\sqrt{2\pi}H_{\rm d})$.

Dust particles in CPDs drift inwards, because they lose their angular momentum by the gas in sub-Keplerian rotation. The radial drift velocity of the particles is \citep{whi72,ada76,wei77}
\begin{equation}
v_{\rm r}=-2\dfrac{\rm St}{\rm St^{2}+1}\eta v_{\rm k},
\label{vr}
\end{equation}
where $v_{\rm k}=r\Omega_{\rm k}$ is the Kepler velocity, and
\begin{equation}
\eta=-\dfrac{1}{2}\left(\dfrac{H_{\rm g}}{r}\right)^{2}\dfrac{\partial \ln{\rho_{\rm g,mid}c_{\rm s}^{2}}}{\partial \ln{r}}
\label{eta}
\end{equation}
is the ratio of the pressure gradient force to the gravity of the central planet.

The collision velocity between the particles is,
\begin{equation}
\Delta v_{\rm dd}=\sqrt{\Delta v_{\rm B}^{2}+\Delta v_{\rm r}^{2}+\Delta v_{\rm \phi}^{2}+\Delta v_{\rm z}^{2}+\Delta v_{\rm t}^{2}},
\label{vdd}
\end{equation}
where $\Delta v_{\rm B}$, $\Delta v_{\rm r}$, $\Delta v_{\rm \phi}$, $\Delta v_{\rm z}$, and $\Delta v_{\rm t}$ are the relative velocities induced by their Brownian motion, radial drift, azimuthal drift, vertical sedimentation, and turbulence, respectively \citep{oku12}. These velocities are $\Delta v_{\rm B}=\sqrt{16k_{\rm B}T/(\pi m_{\rm d})}$, $\Delta v_{\rm r}=|v_{\rm r}({\rm St}_{1})-v_{\rm r}({\rm St}_{2})|$, where ${\rm St}_{1}={\rm St}$ and ${\rm St}_{2}=0.5{\rm St}$, $\Delta v_{\rm \phi}=|v_{\rm \phi}({\rm St}_{1})-v_{\rm \phi}({\rm St}_{2})|$, where $v_{\rm \phi}=-\eta v_{\rm K}/(1+{\rm St}^{2})$, and $\Delta v_{\rm z}=|v_{\rm z}({\rm St}_{1})-v_{\rm z}({\rm St}_{2})|$, where $v_{\rm z}=-\Omega_{\rm K}{\rm St}H_{\rm d,diff}/(1+{\rm St})$ (see \citet{sat16} for the details). The relative velocity induced by turbulence (diffusion) is \citep{orm07}
\begin{equation}
\Delta \varv_{\rm t}=
\begin{cases}
\sqrt{\alpha_{\rm diff}}c_{\rm s}{\rm Re}_{\rm t}^{1/4}\left|{\rm St_{1}}-{\rm St_{2}}\right|, & {\rm St_{1}}\ll {\rm Re}_{\rm t}^{-1/2}, \\
\sqrt{3\alpha_{\rm diff}}c_{\rm s}{\rm St}_{1}^{1/2}, & {\rm Re}_{\rm t}^{-1/2} \ll {\rm St_{1}}\ll 1, \\
\sqrt{\alpha_{\rm diff}}c_{\rm s}\left(\dfrac{1}{1+{\rm St_{1}}}+\dfrac{1}{1+{\rm St_{2}}}\right)^{1/2}, & 1\ll {\rm St_{1}},
\end{cases}
\label{vt}
\end{equation}
where the turbulence Reynolds number is ${\rm Re_{t}}=\nu/\nu_{\rm mol}$. The molecular viscosity is $\nu_{\rm mol}=v_{\rm th}\lambda_{\rm mfp}/2$. We also calculate the dust-to-gas relative velocity, $\Delta v_{\rm dg}$, by setting ${\rm St_{1}}={\rm St}$ and ${\rm St_{2}}\rightarrow0$ in the above equations.

When the collision velocity, $\Delta v_{\rm dd}$, is high, the colliding particles break up rather than merge. The sticking efficiency for a single collision is written as,
\begin{equation}
\epsilon_{\rm grow}=\min\left\{1, -\dfrac{\ln{(\Delta v_{\rm dd}/v_{\rm cr})}}{\ln{5}}\right\},
\label{stfrag}
\end{equation}
from the fitting of the simulations \citep{oku16}. The critical velocity of the fragmentation, which is about $1-50~{\rm m~s^{-1}}$, has been investigated by both experiments and numerical simulations but the exact value is still controversial \citep[e.g.][]{blu00,wad13}. The critical velocity of the icy particles is higher than that of the rocky particles in most of the previous works.

We define the snowline as the orbit where the equilibrium vapour pressure of water, $P_{\rm ev,H_{2}O}$ is equal to its partial pressure, $P_{\rm H_{2}O}$. By the Arrhenius form,
\begin{equation}
P_{\rm ev,H_{2}O}=\exp{\left(-\dfrac{L_{\rm H_{2}O}}{T}+A_{\rm H_{2}O}\right)}~{\rm dyn~cm^{-2}},
\label{PevH2O}
\end{equation}
where $L_{\rm H_{2}O}=6070~{\rm K}$ is the heat of the sublimation of water, and $A_{\rm H_{2}O}=30.86$ is a dimensionless constant \citep{bau97}. Assuming that the gas disc is well mixed in the vertical direction, the partial pressure of water can be expressed as
\begin{equation}
P_{\rm H_{2}O}= \dfrac{\Sigma_{\rm d,H_{2}O}}{\sqrt{2\pi}H_{\rm g}}\dfrac{k_{\rm B}T_{\rm mid}}{\mu_{\rm H_{2}O}},
\label{PH2O}
\end{equation}
where the surface density of water ice is assumed as $\Sigma_{\rm d,H_{2}O}=0.5\Sigma_{\rm d}$ (outside the snowline), and the molecular mass of water is $\mu_{\rm H_{2}O}=18.02m_{\rm H}$.

\subsection{Continuum emission from dust in CPDs}
\label{emission}
The radial distribution of the peak mass of the dust at each distance from the planet is calculated by the way explained in Section \ref{evolution}. We then calculate the continuum emission from the dust. The emission depends on the size of the particles; therefore we redistribute the mass of the particles at each orbital place with an assumed size frequency distribution (SFD). We assume that the number of the particles with the radii of $a$ to $a+{\rm d}a$ at the orbit is proportional to $a^{-q}$. In this case, the surface density of the particles with the size of $a$ (to $a+{\rm d}a$) is 
\begin{equation}
\Sigma_{\rm d,a}(a)=\Sigma_{\rm d,0}a^{3-q},
\label{sfd}
\end{equation}
where
\begin{equation}
\Sigma_{\rm d,0}=\dfrac{(4-q)\Sigma_{\rm d}}{R_{\rm d}^{4-q}-a_{\rm min}^{4-q}}.
\label{Sigma_dzero}
\end{equation}
We assume the minimum size of the particles as $a_{\rm min}=0.1~{\rm\mu m}$.

The vertical optical depth of the disc for the wavelength of $\lambda(=c/\nu)$ is,
\begin{equation}
\tau_{\nu}=\int^{R_{\rm d}}_{a_{\rm min}}\Sigma_{{\rm d},a}\kappa_{\rm abs}~da,
\label{tau}
\end{equation}
where $\kappa_{\rm abs}$ is the absorption mass opacity for the wavelength of $\lambda$ by the particles with the size of $a$. Here, we ignore the scattering opacity. The absorption opacity is,
\begin{equation}
\kappa_{\rm abs}=\dfrac{3}{4a}\dfrac{1}{\rho_{\rm int,opa}}Q_{\rm abs},
\label{kappa_abs}
\end{equation}
where $Q_{\rm abs}$ is the dimensionless absorption coefficient, and $\rho_{\rm int,opa}=1.675~{\rm g~cm^{-3}}$ is the internal density for the calculations of the opacity in both sides of the snowline \citep{bir18}\footnote{We distinguish $\rho_{\rm int,opa}$ from $\rho_{\rm int}$ to obtain consistency with that we use a single composition dust model proposed in \citet{bir18} for the value of the refractive index in Eqs. (\ref{qabs1}) and (\ref{qabs2}).}. We use a model of the coefficient proposed in \citet{kat14},
\begin{equation}
Q_{\rm abs}=
\begin{cases}
Q_{\rm abs,1}, & 2\pi a/\lambda\leq1, \\
\min(Q_{\rm abs,2},Q_{\rm abs,3}), & 2\pi a/\lambda>1.
\end{cases}
\label{qabs}
\end{equation}
When the dust is much smaller than the wave length, in other words $2\pi a/\lambda\ll1$, the opacity goes into the Rayleigh regime. The coefficient is approximated as
\begin{equation}
Q_{\rm abs}\simeq Q_{\rm abs,1}\equiv\dfrac{24nk}{(n^{2}+2)^{2}}\dfrac{2\pi a}{\lambda},
\label{qabs1}
\end{equation}
where $n$ and $k$ are the real and imaginary parts of the refractive index, which depend on the wavelength and the composition of the dust. For the values of $n$ and $k$, we use the `DSHARP dust' opacity model proposed in \citet{bir18} (see Figure 2 of the paper) instead of the model of \citet{kat14}. The opacity could change by a factor of 2 to 3 depending on the dust opacity models. When the dust is much larger than the wavelength (i.e. $2\pi a/\lambda\gg1$), the opacity goes into the geometric optics regime. In optically thin cases, the coefficient can be approximated as
\begin{equation}
Q_{\rm abs}\simeq Q_{\rm abs,2}\equiv\dfrac{8k}{3n}\dfrac{2\pi a}{\lambda}\{n^{3}-(n^{2}-1)^{3/2}\},
\label{qabs2}
\end{equation}
and in optically thick cases, we set the coefficient as $Q_{\rm abs}\simeq Q_{\rm abs,3}=0.9$ (see \citet{kat14} for the details).

The total continuum emission from the dust in the CPD is then,
\begin{equation}
F_{\rm emit}=\dfrac{2\pi\cos{i}}{d^{2}}\int^{r_{\rm out}}_{r_{\rm in}}\left\{1-\exp\left(-\dfrac{\tau_{\nu}}{\cos{i}}\right)\right\}B_{\nu}rdr,
\label{Femit}
\end{equation}
where $i$ and $d$ are the inclination of the CPD, which is assumed to be the same with that of the parental PPD, and the distance to the CPD from Earth, respectively \citep{kep19}. The Planck function, $B_{\nu}$, depends on the temperature of the dust, which is assumed to be the same with the midplane disc temperature, $T$, because the CPDs are optically thin (see Section \ref{evolution})\footnote{The dust temperature should be assumed to be the same with the disc temperature at the height where $\tau_{\nu}=1$ when the disc is optically thick.}. In the steady dust evolution cases, we consider the situations that the dust particles do not drift outwards and exist only inside $r_{\rm inf}$, which makes $r_{\rm out}$ possible to be replaced to $r_{\rm inf}$.

Almost all of the supplied dust grows large to the pebble size and drifts inwards. We note that, however, there is a possibility that the supplied dust goes to the outside of the modelled region ($1/3~R_{\rm H}<r<R_{\rm H}$; see Fig. \ref{fig:models}) by the gas outflow on the midplane (if it exists) or their diffusion, which can not be reproduced by our current model. However, the gas density outside the region we modelled is $\sim100$ times smaller than that of the inside ($r<1/3~R_{\rm H}$) in one of the recent numerical simulations \citep{sch20}. Thus, if the dust-to-gas density ratio is same in the both inside and outside, although the total surface area of the outside is $\sim10$ times larger than that of the inside, the total dust emission from the outside should be $\sim10$ times smaller than that from the inside when the disc is optically thin. Therefore, we consider that the emission from the outside is negligible.

\subsection{Parameter settings} \label{parameters}
We summarise the parameters used in this work in Tab. \ref{tab:parameters}. We change the value of the two important planet properties, the planet mass and the gas accretion rate, in the calculations for each planet. We also change two properties of the CPDs, the strength of the $\alpha$-turbulence in the CPDs, and the dust-to-gas mass ratio in the inflow, for each planet-CPD system. The estimates of the two planet properties by previous works are summarised in Tab. \ref{tab:estimates} (see Section \ref{constraints} for the detailed explanations). We investigate the dependence of the flux density of dust emission from the CPDs on the properties in Section \ref{planet-properties} and find constraints on the properties in Section \ref{constraints}. The other parameters are fixed; we show that they have only little impacts on the results in Section \ref{properties}.

\begin{table*}[htbp]
\caption{Parameters. The four varied quantities are in boldface.}
\label{tab:parameters}
\centering
\small
\begin{tabular}{llll}
\hline
Description & Symbol & Value & Reference \\
\hline\hline
\multicolumn{4}{l}{General} \\
\hline
{\bf Dust-to-gas mass ratio of gas inflow} & $x$ & $0.0001, 0.001, 0.01$ & - \\
{\bf Strength of turbulence in CPD} & $\alpha$ & $10^{-5}, 10^{-3}$ & - \\
Mass fraction of helium in gas & $Y$ & $0.27$ & - \\
Heat of water sublimation & $L_{\rm H_{2}O}$ & $6070~{\rm K}$ & \citet{bau97} \\
Constant of water vapour pressure & $A_{\rm H_{2}O}$ & $30.86$ & \citet{bau97} \\
Icy dust fragmentation speed & $v_{\rm ice}$ & $50~{\rm m~s^{-1}}$ & \citet{wad13} \\
Rocky dust fragmentation speed & $v_{\rm rock}$ & $5~{\rm m~s^{-1}}$ & \citet{wad13} \\
Icy dust internal density & $\rho_{\rm int,ice}$ & $1.4~{\rm g~cm^{-3}}$ & - \\
Rocky dust internal density & $\rho_{\rm int,rock}$ & $3.0~{\rm g~cm^{-3}}$ & - \\
Size (radius) of dust in gas inflow & $R_{\rm d,0}$ & $1~\mu{\rm m}$ & \citet{bae19} \\
Minimum size (radius) of dust particles & $a_{\rm min}$ & $0.1~\mu{\rm m}$ & - \\
(Minus) power-low index of SFD of dust & $q$ & $3.5$ & - \\
Real part of refractive index & $n$ & 2.298 & \citet{bir18} \\
Imaginary part of refractive index & $k$ & 0.02146 & \citet{bir18} \\
Dust internal density (for opacity calc.) & $\rho_{\rm int,opa}$ & $1.675~{\rm g~cm^{-3}}$ & \citet{bir18} \\
Global radiation efficiency of gas accretion & $\eta_{\rm eff}$ & $0.95$ & \citet{mar19} \\
Dynamo region to surface mean magnetic field ratio & $f_{\rm surf}$ & $3.5$ & \citet{chr09} \\
Ratio of ohmic to total dissipation & $f_{\rm ohm}$ & $1$ & \citet{chr09} \\
Constant of proportionality (for magnetic field calc.) & $c$ & $0.63$ & \citet{chr09} \\
Efficiency factor (for magnetic field calc.) & $F$ & $1$ & \citet{chr09} \\
Aspect ratio of CPD (for magnetic field calc.) & $h_{\rm asp}$ & $0.1$ & \citet{chr09} \\
\hline
\multicolumn{4}{l}{PDS~70 system} \\
\hline
Host star mass & $M_{\rm star}$ & $0.76~M_{\odot}$ & \citet{mul18} \\
Distance from Earth & $d$ & $113.43~{\rm pc}$ & \citet{gaia18} \\
Inclination of CPDs (same with PPD) & $i$ & $51.7^{\circ}$ & \citet{kep19} \\
\hline
\multicolumn{4}{l}{PDS~70~b} \\
\hline
{\bf Planet mass} & $M_{\rm p}$ & $(0.5-20)~M_{\rm J}$ & - \\
{\bf Gas accretion rate} & $\dot{M}_{\rm g,pl}$ & $(10^{-9}-10^{-6.5})~M_{\rm J}~{\rm yr^{-1}}$ & - \\
Semimajor axis & $a_{\rm pl}$ & $20.6~{\rm au}$ & \citet{haf19} \\
Radius & $R_{\rm pl}$ & $2.0~R_{\rm J}$ & \citet{wan21} \\
Effective temperature & $T_{\rm pl,eff}$ & $1392~{\rm K}$ & \citet{wan21} \\
Temperature of PPD & $T_{\rm irr,PPD}$ & $51~{\rm K}$ & \citet{law24} \\
\hline
\multicolumn{4}{l}{PDS~70~c} \\
\hline
{\bf Planet mass} & $M_{\rm p}$ & $(0.5-20)~M_{\rm J}$ & - \\
{\bf Gas accretion rate} & $\dot{M}_{\rm g,pl}$ & $(10^{-9}-10^{-6.5})~M_{\rm J}~{\rm yr^{-1}}$ & - \\
Semimajor axis & $a_{\rm pl}$ & $34.5~{\rm au}$ & \citet{haf19} \\
Radius & $R_{\rm pl}$ & $2.0~R_{\rm J}$ & \citet{wan21} \\
Effective temperature & $T_{\rm pl,eff}$ & $1051~{\rm K}$ &  \citet{wan21} \\
Temperature of PPD & $T_{\rm irr,PPD}$ & $32~{\rm K}$ & \citet{law24} \\
\hline
\end{tabular}
\end{table*}

\begin{table*}[htbp]
\caption{Estimates of important properties by previous works.}
\label{tab:estimates}
\centering
\small
\begin{threeparttable}[h]
\begin{tabular}{llll}
\hline
PDS~70~b & Value & Observation types & Reference \\
\hline
Planet mass [$M_{\rm J}$] & $5-9$ & IR colours and evolution model & \citet{kep18} \\
& $2-17$ & SED (IR)\tnote{1} & \citet{mul18} \\
& $12$ & H$\alpha$ 10\% and 50\% width & \citet{aoy19} \\
& $1.1-11.6$ & Dynamical stability (95\%) & \citet{wan21} \\
& $4$ & Gap depth & \citet{por23} \\
\hline
Gas accretion rate [$M_{\rm J}~{\rm yr^{-1}}$] & $1\times10^{-8\pm1}$ & SED (IR) and TTS empirical relation\tnote{1} & \citet{wag18} \\
& $2\times10^{-8\pm0.4}$ & H$\alpha$ 10\% width and TTS empirical relation & \citet{haf19} \\
& $1\times10^{-8\pm0.6}$ & H$\alpha$ luminosity and magnetospheric accretion\tnote{1,2} & \citet{tha19} \\
& $4\times10^{-8}$ & H$\alpha$ 10\% and 50\% width and H$\alpha$ luminosity\tnote{1}  & \citet{aoy19} \\
\hline
MMdot [$M_{\rm J}^{2}~{\rm yr^{-1}}$] & $<1.26\times10^{-6}$ & Br$\gamma$ luminosity and TTS empirical relation\tnote{1} & \citet{chr19a} \\
& $10^{-6.8}-10^{-6.3}$ & SED (IR) and CPD model\tnote{1} & \citet{chr19b} \\
& $4.8\times10^{-7}$ & H$\alpha$ luminosity\tnote{1} & \citet{aoy19} \\
& $(2.5-7.5)\times10^{-7}$ & SED (IR)\tnote{1} & \citet{sto20} \\
& $(1-10)\times10^{-7}$ & SED (IR) and CPD model & \citet{wan21} \\
& $(1.6\pm0.23)\times10^{-8}$ & UV and H$\alpha$ luminosity\tnote{1} & \citet{zho21} \\
\hline
PDS~70~c & & & \\
\hline
Planet mass & $4-12$ & K-L colour and evolution model & \citet{haf19} \\
& $10$ & H$\alpha$ 10\% and 50\% width & \citet{aoy19} \\
& $1.4-14.5$ & Dynamical stability (95\%) & \citet{wan21} \\
& $4$ & Gap depth & \citet{por23} \\
\hline
Gas accretion rate & $1\times10^{-8\pm0.4}$ & H$\alpha$ 10\% width and TTS empirical relation\tnote{1} & \citet{haf19} \\
& $1\times10^{-8.1\pm0.6}$ & H$\alpha$ luminosity and magnetospheric accretion\tnote{1,2} & \citet{tha19} \\
& $1\times10^{-8}$ & H$\alpha$ 10\% and 50\% width and H$\alpha$ luminosity\tnote{1} & \citet{aoy19} \\
\hline
MMdot & $1\times10^{-7}$ & H$\alpha$ luminosity\tnote{1} & \citet{aoy19} \\
& $(1-10)\times10^{-7}$ & SED (IR) and CPD model & \citet{wan21} \\
\hline
PDS~70 & & & \\
\hline
Gas accretion rate & $(1.4\pm0.8)\times10^{-7}$ & H$\alpha$ profiles and magnetospheric accretion & \citet{tha20} \\
\hline
\end{tabular}
\begin{tablenotes}
\item[1] The effects of the extinction is not included.
\item[2] The planet mass is assumed as $6~M_{\rm J}$.
\end{tablenotes}
\end{threeparttable}
\end{table*}

\section{Results} \label{results}
\subsection{Distribution of gas and dust in the CPD of PDS~70~c} \label{distribution}
We first investigate the detailed evolution of the dust in the CPD of PDS~70~c and the continuum emission from the evolving dust. Here, we show the case where $M_{\rm pl}=10~M_{\rm J}$ and $\dot{M}_{\rm g,pl}=2\times10^{-7}~M_{\rm J}~{\rm yr}^{-1}$ (same with the `plausible case' obtained in Section \ref{constraints-PDS70c}) with $x=0.01$ and $\alpha=10^{-4}$ as the fiducial case. The angular momentum bias of the gas inflow is then $l=0.57$ (Eq. (\ref{l})). We then change the planet and disc properties from this plausible case and investigate the effects of each change.

\subsubsection{Structures of gas and temperature in the CPD} \label{gastemp}
\begin{figure}[tbp]
\centering
\includegraphics[width=0.95\linewidth]{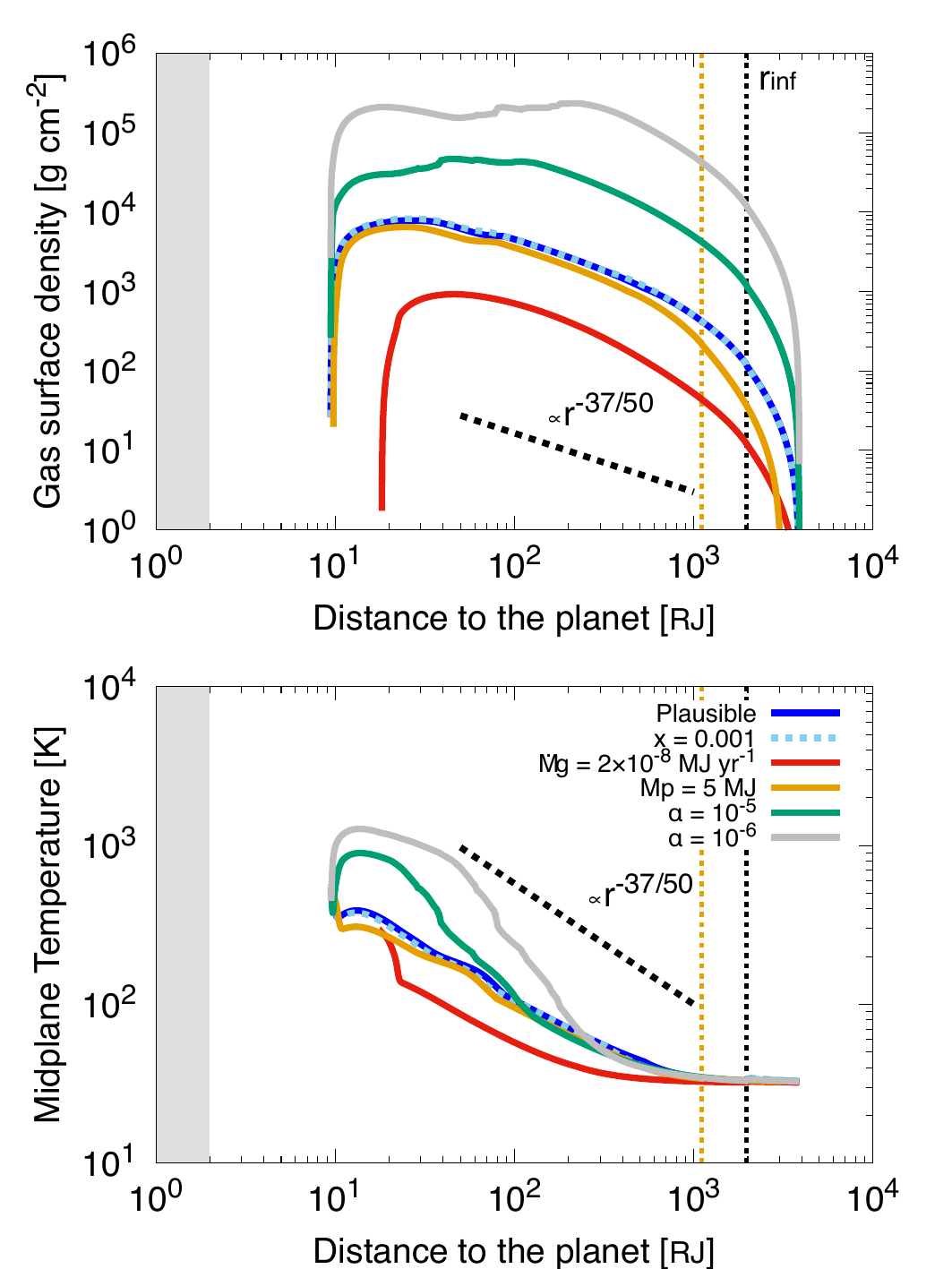}
\caption{Gas surface density and disc midplane temperature of the CPD of PDS~70~c. The blue curves represent the plausible case: $M_{\rm pl}=10~M_{\rm J}$, $\dot{M}_{\rm g,pl}=2\times10^{-7}~M_{\rm J}~{\rm yr}^{-1}$, $x=0.01$, and $\alpha=10^{-4}$. The sky-blue, red, orange, green, and grey curves represent the cases with $x=0.001$, $\dot{M}_{\rm g,pl}=2\times10^{-8}~M_{\rm J}~{\rm yr}^{-1}$, $M_{\rm pl}=5~M_{\rm J}$, $\alpha=10^{-5}$, and $\alpha=10^{-6}$, respectively. The vertical dotted lines are the outer edges of the gas inflow regions, $r=r_{\rm inf}$, which only depends on $M_{\rm pl}$. The oblique dashed lines in the upper and lower panels represent the slopes of $\Sigma_{\rm g}\propto r^{-37/50}$ and $T\propto r^{-19/25}$, respectively. Shaded grey regions represent the planetary atmosphere.}
\label{fig:gas}
\end{figure}

Figure \ref{fig:gas} shows the gas surface density and the midplane temperature of the CPD. The gas disc is truncated at $r_{\rm in}$ and $r_{\rm out}$. The gas surface density of the plausible case (blue curves) is about $10^{3}-10^{4}~{\rm g~cm^{-2}}$ inside the gas inflow region, $r\leq r_{\rm inf}$, where $r_{\rm inf}$ is expressed as the vertical dotted lines. The slopes of the gas surface density and the midplane temperature inside $r_{\rm inf}$ ($100\lesssim r\lesssim1000~R_{\rm J}$) are close to $\Sigma_{\rm g}\propto r^{-37/50}$ and $T_{\rm mid}\propto r^{-19/25}$ (oblique dashed lines), which are derived as follows. The gas surface density can be approximated as $\Sigma_{\rm g}\propto r^{-3/2}T_{\rm mid}^{-1}$, where $\dot{M}_{\rm g,pl}$ and $\alpha$ are uniform. Thus, when $\tau_{\rm R}=\kappa_{\rm R}\Sigma_{\rm g}\ll1$ and the midplane temperature is determined by the viscous heating, $T_{\rm mid}\propto r^{-3/4}\tau_{\rm R}^{-1/4}\propto\kappa_{\rm R}^{-1/3}r^{-1/2}$. The opacity is $\kappa_{\rm R}\propto T_{\rm mid}^{2}Z_{\rm \Sigma,est}$ outside the snowline, where we assume $Z_{\rm\Sigma,est}\propto r^{2.3}$ (see Appendix \ref{opacity}). Then, we get $\Sigma_{\rm g}\propto r^{-37/50}$ and $T\propto r^{-19/25}$. The temperate of the outermost region of the CPD ($r\gtrsim1000~R_{\rm J}$) is determined by the PPD temperature, $T_{\rm irr,PPD}=32~{\rm K}$.

When the gas accretion rate is lower (red curves) than the plausible case, the gas surface density and temperature are also lower, which are consistent with Eq. (\ref{SigmagCW}) and (\ref{dotEv}). Also, the position of the disc inner edge is outside that of the plausible case (see Section \ref{inneredge}). When the planet mass is lower (orange curves), the outer edges of the disc ($r_{\rm out}=1/3~R_{\rm H}$) and the gas inflow region ($r_{\rm inf}=25l^{2}/48~R_{\rm H}$), where $R_{\rm H}\propto M_{\rm pl}^{1/3}$, are smaller (see also Eq. (\ref{l}) for the detailed $M_{\rm pl}$ dependence). However, the value of the gas surface density and temperature is almost the same with the plausible case. When the turbulence is weaker (green curves), the gas surface density is higher, which is consistent with Eq. (\ref{SigmagCW}) showing roughly $\Sigma_{\rm g}\propto\alpha^{-1}$. The disc temperature is almost the same with the plausible case ($r\geq100~R_{\rm J}$), because $\Sigma_{\rm g}$ and $\alpha$ in the viscus dissipation rate ($\dot{E}_{\rm v}$) are cancelled out (Eq. (\ref{dotEv})). The steep slopes of the temperature profiles around $30\lesssim r\lesssim100~R_{\rm J}$ are formed by that the gas opacity dominates the opacity of the disc in that region. We also plot the profiles with $\alpha=10^{-6}$ as grey curves, but the disc may be gravitationally unstable in this case (see Section \ref{properties} and Appendix \ref{GI}). 

\subsubsection{Evolution and emission of dust in the CPD} \label{detailed}
\begin{figure*}[htbp]
\centering
\includegraphics[width=0.95\linewidth]{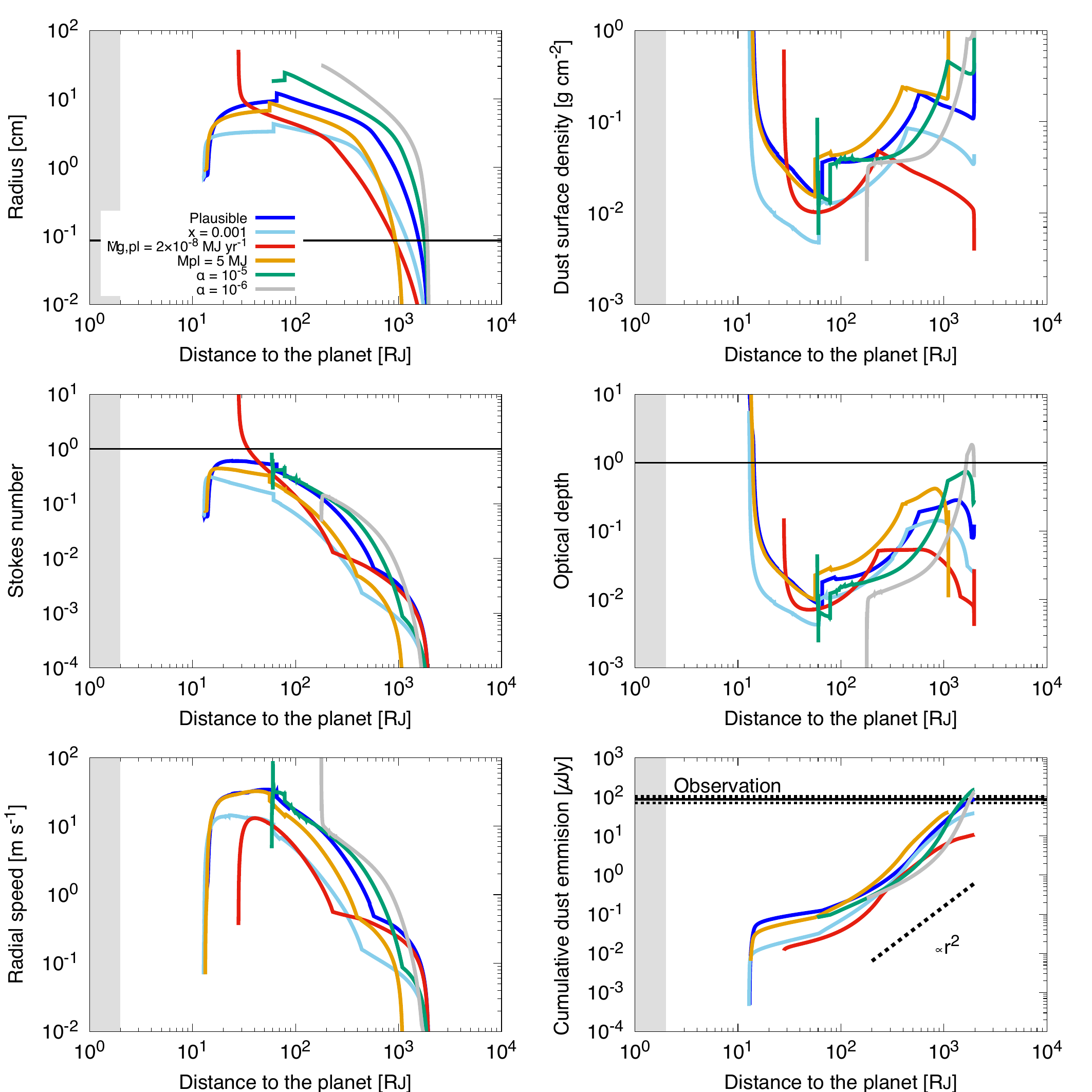}
\caption{Dust evolution in the CPD of PDS~70~c with various parameter sets. The left top, left middle, left bottom, right top, right middle, and right bottom panels represent the dust radial profiles of radius, Stokes number, radial drift speed, surface density, optical depth, cumulative flux density contribution of dust emission from the centre of the disc, respectively. The colour variation is same with Fig. \ref{fig:gas}. The black lines in the left top, left middle, right middle, and right bottom panels are the wavelength of the observation ($\lambda=855~{\rm \mu m}$), unity (i.e. showing the highest radial drift speed), unity (i.e. showing optically thick or thin), and the observed dust emission value $86\pm16~\mu{\rm Jy}$ \citep{ben21}, respectively.}
\label{fig:evolution}
\end{figure*}

We then show the evolution of the dust in the gas disc explained in Section \ref{gastemp}. Figure \ref{fig:evolution} shows the dust evolution in the CPD of PDS~70~c with various sets of parameters. First, we explain the plausible case, $M_{\rm pl}=10~M_{\rm J}$ and $\dot{M}_{\rm g,pl}=2\times10^{-7}~M_{\rm J}~{\rm yr}^{-1}$ with $x=0.01$ and $\alpha=10^{-4}$ (blue curves). The left top panel shows that the small dust particles grow quickly to cm-sized particles (i.e. pebbles) by mutual collision at the place where they are supplied to the CPD ($r=r_{\rm inf}$) and then drift towards the central planet. When the growth timescale becomes longer than the drift timescale, the dust drift starts. As a result, the dust radius (of the peak mass) is larger than the observed wave length, $\lambda=855~\mu{\rm m}$. This picture of evolution of dust is the same with the one in the CPD of Jupiter \citep{shi17}. The left middle panel shows that the Stokes number of dust also increases as the particles drift inwards (see Eq. (\ref{St})). The changes of the slopes from gradual to steep are formed when the dust goes into the Stokes regime from the Epstein regime. The dust particles do not grow so much once they start to drift, and the surface density of the drifting dust is roughly uniform or gradually larger as $r$ is larger (right top panel). The optical depth is also almost uniform or gradually larger as $r$ is larger, and the disc is optically thin in the whole region due to the radial drift of the particles (right middle). As a result, the dust emission per unit area is almost uniform, resulting in that the slopes of the cumulative dust emission are about $\propto r^{2}$ (right bottom).

The small steps of the profiles around $50-80~R_{\rm J}$ are formed by the snowline. The size of the particles inside the snowline is determined by fragmentation and by radial drift outside the snowline, which is also shown in the Jovian CPD case \citep{shi23}. The stronger turbulence causes efficient fragmentation at the inner region, resulting in a smaller radius of dust.

The right bottom panel of Fig. \ref{fig:evolution} shows that there is a positive correlation between the dust-to-gas mass ratio of the inflow and the total flux density of dust emission (blue and sky-blue curves). However, it is weaker than the dependence on the other properties such as $M_{\rm pl}$ and $\dot{M}_{\rm g,pl}$ (see also Section \ref{planet-properties}), which can be explained as follows. As the dust-to-gas density ratio in the inflow is small (sky-blue), in other words, as the dust mass flux onto the CPD is small, the collisional growth is less efficient, because there are less dust particles (i.e. lower $\rho_{\rm d,mid}$). Then, the timescale of dust is longer, and so the dust particles grow smaller before they start to drift (left top and middle panels). However, that means the radial speed ($|v_{\rm r}|$) is also slower (left bottom), resulting in the effect of the high $\dot{M}_{\rm d}$ to the dust surface density being almost cancelled out (Eq. (\ref{Mdotdust})). As a result, the right top panel shows that $x$ dependence of $\Sigma_{\rm d}$ is weak. Also, the dust size is small (i.e. closer to the wavelength) when $x$ is small (left top), making the $x$ dependence of $\tau_{\lambda}$ even weaker (right middle), because the opacity is $\kappa_{\rm abs}\propto a^{-1}$ when the dust size is larger than the wavelength (see Section \ref{emission}). In total, the $x$ dependence of the flux density of dust emission is relatively weak.

When the gas accretion rate is lower than the plausible case and the dust-to-gas mass ratio in the gas inflow is fixed (red curves), the dust mass accretion rate is also lower. Then, the growth timescale of dust just supplied to the discs is longer. Also, when the gas accretion rate is low, the gas surface density is low (upper panel of Fig. \ref{fig:gas}). Then, the Stokes number is almost the same with the plausible case even with smaller dust (left middle). Therefore, considering the mass conservation ($\Sigma_{\rm d}\propto\dot{M}_{\rm d}/|v_{\rm r}|$) and the fixed $x\equiv\dot{M}_{\rm d,tot}/\dot{M}_{\rm g,tot}\approx\dot{M}_{\rm d}/\dot{M}_{\rm g,pl}$, $\Sigma_{\rm d}$ is about $\Sigma_{\rm d}\propto\dot{M}_{\rm g,pl}$. Actually, the dust surface density at the outer region of the disc is lower than the plausible case (right top). Then, the dust emission from the outer region (dominating the total flux density) is smaller, making the total flux density of dust mission smaller as well (right bottom). When the effect of dust size to the dust emission flux density is negligible, the dependence can be approximated as $F_{\rm emit}\propto\dot{M}_{\rm g,pl}$, which is shown in Section \ref{planet-properties}. Also, the orbital position where the dust goes into the Stokes regime from the Epstein regime (around $200~R_{\rm J}$) is inner than that of the plausible case (around $600~R_{\rm J}$) due to the lower gas surface density, resulting in the Stokes number smaller around $50-600~R_{\rm J}$ (left middle).

The dust emission flux density also depends on the planet mass (orange curves). First, the surface area of the dust existing region is $\pi r_{\rm inf}^{2}\propto(l^{2}R_{\rm H})^{2}$. Here, $R_{\rm B}\gg R_{\rm H}$, meaning that roughly $l^{2}R_{\rm H}\propto R_{\rm B}\propto M_{\rm pl}$. Second, the dust surface density is $\Sigma_{\rm d}\propto|v_{\rm r}|^{-1}\propto ({\rm St}\Omega_{\rm K})^{-1}$. The left lower panel of Fig. \ref{fig:evolution} shows that the Stokes number is small when the planet mass is small (about ${\rm St}\propto M_{\rm pl}^{1/2}$), because the dust starts to grow at a shorter orbit when the planet mass is smaller ($r_{\rm inf}\propto M_{\rm pl}$). Therefore, when ${\rm St}\propto M_{\rm pl}^{1/2}$, $\Sigma_{\rm d}\propto M_{\rm pl}^{-1}$. Finally, the figure shows that the dust size dependence of the flux density of dust emission is weak. In conclusion, $F_{\rm emit}\propto r_{\rm inf}^{2}\times\Sigma_{\rm d}\propto M_{\rm pl}$, which is shown in Section \ref{planet-properties}.

When the turbulence is weak (green curves), the dust inflow rate is the same but the gas surface density of the disc is high (upper panel of Fig. \ref{fig:gas}). Thus, the Stokes number is small at the outer part of the disc ($r\gtrsim1000~R_{\rm J}$ in the left middle panel of Fig. \ref{fig:evolution}; Epstein regime), resulting in slow radial drift of the dust particles (left bottom). As a result, the dust surface density is large (right top), which makes the total dust emission large (right bottom).

\subsection{Effects of planet properties} \label{planet-properties}
We then investigate the impacts of the three fundamental properties of the planet, the planet mass ($M_{\rm pl}$), the gas accretion rate to the planet ($\dot{M}_{\rm g,pl}$), and their product called `MMdot' ($M_{\rm pl}\dot{M}_{\rm g,pl}$), on the flux density of dust emission from the CPDs. These properties are also estimated by other observations. For example, a fitting of the spectral energy distribution (SED) of the planet can suggest the planet mass by an atmospheric model \citep{mul18}. The orbital stability also constraints the planet mass \citep{wan21}. The planet mass can also be estimated from the width and depth of the gap \citep{duf15a,kan16,por23}. The gas accretion rate and MMdot are also important. The gas accretion rate can be estimated by the band width of H$\alpha$ emission line \citep{aoy19,haf19}. Also, the luminosity of H$\alpha$ emission and the SED of infrared (IR) provide estimates of the MMdot \citep{zhu15a,wag18,aoy19,wan21}. Due to the much wider range of the gas accretion rate (two to three order of magnitude) than that of the planet mass (only one order of magnitude), the flux of such emission can constraint mainly the gas accretion rate.

\begin{figure*}[htbp]
\centering
\includegraphics[width=0.32\linewidth]{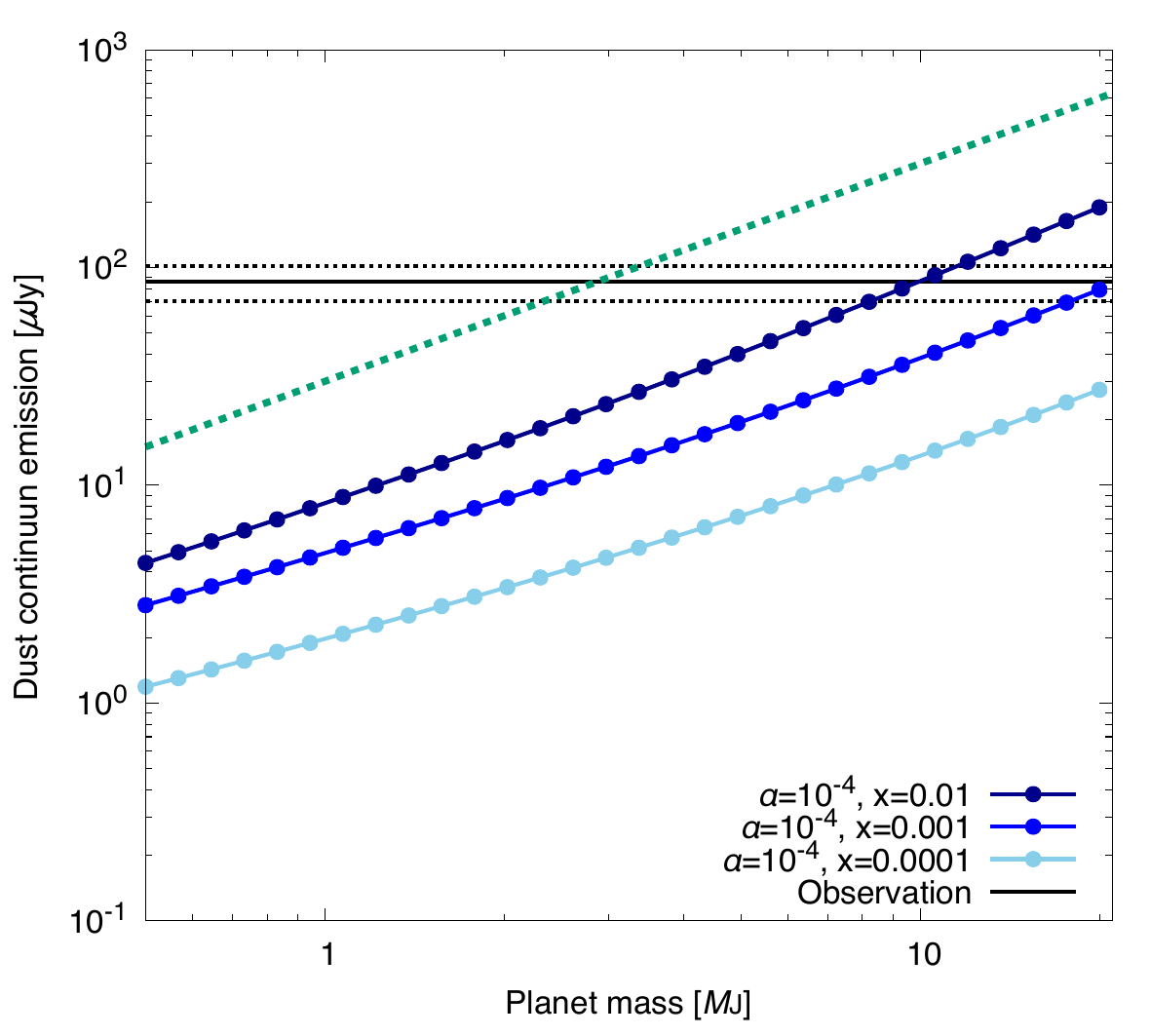}
\includegraphics[width=0.32\linewidth]{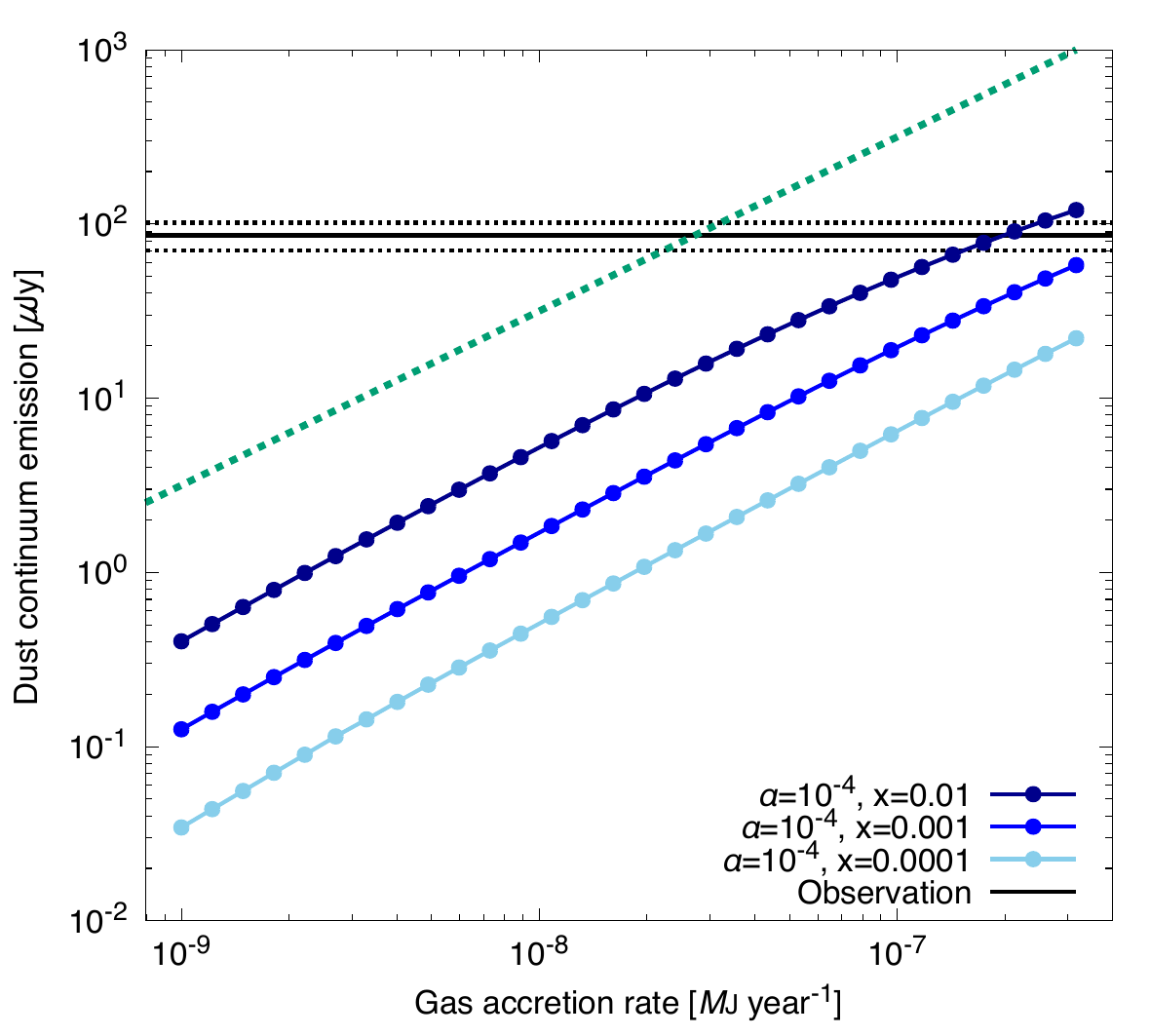}
\includegraphics[width=0.32\linewidth]{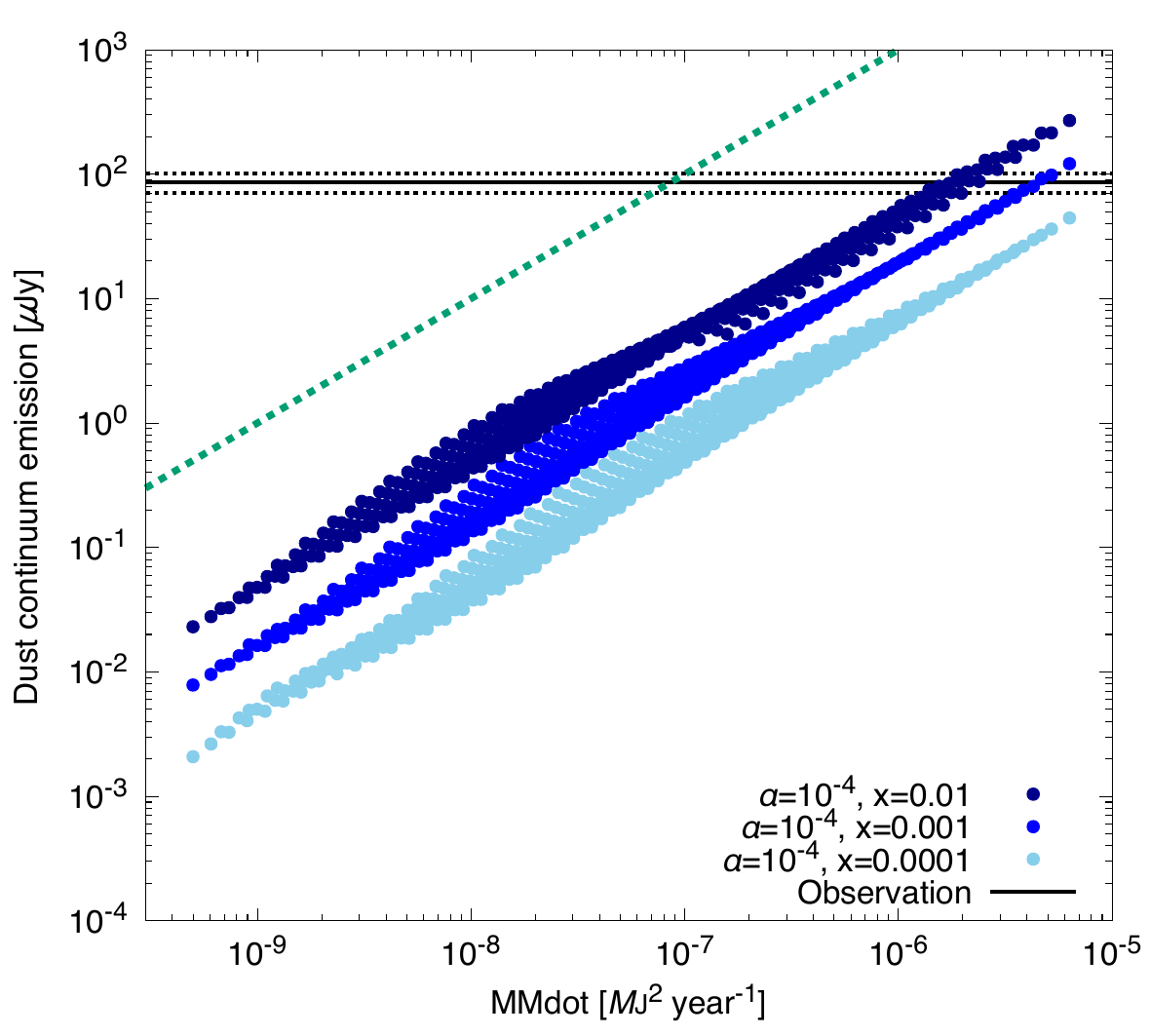}
\caption{Dust continuum emission from the CPDs of PDS~70~c. The dark-blue, blue, and sky-blue curves represent the results when the dust-to-gas mass ratio in the gas inflow is $x=0.01$, $0.001$, and $0.0001$, respectively. The strength of turbulence in the CPDs is fixed as $\alpha=10^{-4}$. The horizontal lines represent the observed value, $86\pm16~\mu{\rm Jy}$ \citep{ben21}. The left, central, and right panels represent the dependence on the planet mass, gas accretion rate, and MMdot, respectively. (Left) The gas accretion rate is fixed as $\dot{M}_{\rm g,pl}=2\times10^{-7}~M_{\rm J}~{\rm yr}^{-1}$. The green dotted lines represent the slopes of $F_{\rm emit}\propto\dot{M}_{\rm g,pl}$. (Centre) The planet mass is fixed as $M_{\rm pl}=10~M_{\rm J}$. The green lines are $F_{\rm emit}\propto M_{\rm pl}$. (Right) Both of the planet mass and the gas accretion rate are changed. The green lines are $F_{\rm emit}\propto M_{\rm pl}\dot{M}_{\rm g,pl}$.}
\label{fig:planet-properties}
\end{figure*}

Figure \ref{fig:planet-properties} represents the dependence of the dust emission from the CPD of PDS~70~c on the three properties. The black lines are the observed value of the dust emission from PDS~70~c. Here, we focus on the dependence, and we discuss the constraints on the properties obtained from the observations in Section \ref{constraints}.

The left panel represents the planet mass dependence when the gas accretion rate is fixed as $\dot{M}_{\rm g}=2\times10^{-7}~M_{\rm J}~{\rm yr}^{-1}$. There is a positive correlation between the planet mass and the total continuum emission. The dependence is about $F_{\rm emit}\propto M_{\rm pl}$ (green dotted lines). If the flux density is in proportion to the surface area of the dust existing region of the CPDs, $F_{\rm emit}\propto r_{\rm inf}^{2}\propto(l^{2}R_{\rm H})^{2}\propto R_{\rm B}^{2}\propto M_{\rm pl}^{2}$, because $R_{\rm B}\gg R_{\rm H}$ (see Eq.(\ref{l})). However, the dust is supplied at $r_{\rm inf}$, meaning the dust can grow larger when the planet mass is large, witch makes the dust surface density lower because of the faster dust drift. In total, we get $F_{\rm emit}\propto M_{\rm pl}$ when the dust size dependence of the flux density of dust emission is weak (see Section \ref{detailed} for the detailed explanation).

The central panel represents the gas accretion rate dependence when $M_{\rm pl}=10~M_{\rm J}$. There is also a clear positive correlation between the gas accretion rate and the total flux density of dust continuum emission. The dependence is about $F_{\rm emit}\propto\dot{M}_{\rm g,pl}$ (green dotted line). The flux density of dust emission is roughly in proportion to $\Sigma_{\rm d}$, where $\Sigma_{\rm d}\propto\dot{M}_{\rm d}\propto\dot{M}_{\rm g,pl}$. We note that the gas surface density depends on the gas accretion rate, and the fluid regimes of dust (Epstein or Stokes regimes) are determined by the gas surface density, which affects the evolution of dust. However, the effects of the difference of the regimes on the total flux density of dust emission almost cancel each other out (see Section \ref{detailed} for the detailed explanation).

The right panel shows the MMdot dependence of the dust emission. As we discussed above, the dust emission is in proportion to the planet mass and the gas accretion rate. Therefore, the MMdot dependence is also about $F_{\rm emit}\propto M_{\rm pl}\dot{M}_{\rm g,pl}$, and it suggests that MMdot is one of the most essential parameters determining the flux density of dust emission from CPDs as well as $M_{\rm pl}$ and $\dot{M}_{\rm g,pl}$. This fact is helpful for the comparisons of the estimates obtained by other types of observation such as H$\alpha$ luminosity and SED of near-infrared.

\subsection{Effects of turbulence in CPDs and other properties} \label{properties}
In this section, we investigate the effects of the parameters other than the planet mass and the gas accretion rate, which are fixed as $M_{\rm pl}=10~M_{\rm J}$ and $\dot{M}_{\rm g,pl}=2\times10^{-7}~M_{\rm J}~{\rm yr}^{-1}$ (the `plausible case' in Section \ref{constraints-PDS70c}).

\begin{figure}[htbp]
\centering
\includegraphics[width=0.99\linewidth]{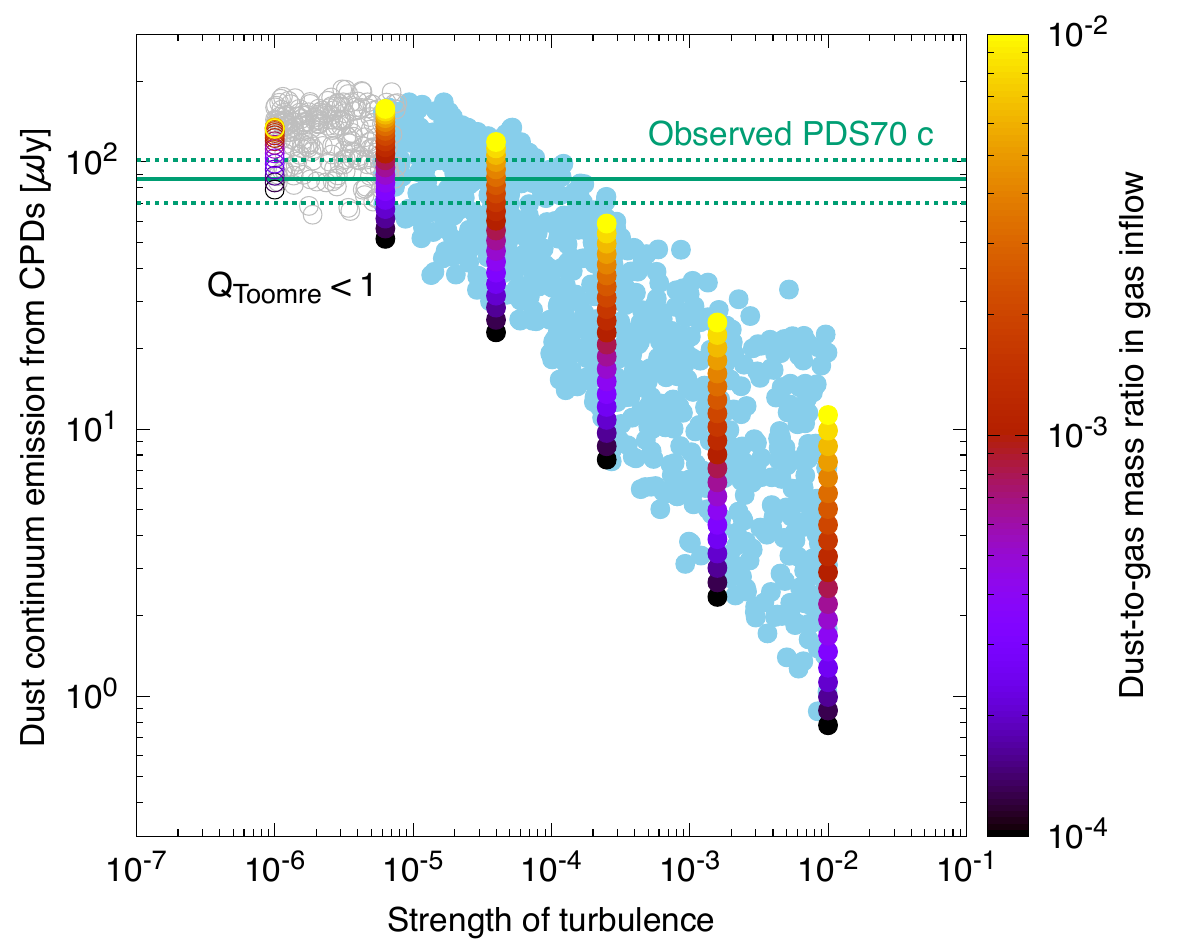}
\caption{Dependence of the dust emission from the CPD of PDS~70~c on the strength of turbulence and the dust-to-gas mass ratio in the inflow (colour). The properties of the planet are fixed as $M_{\rm pl}=10~M_{\rm J}$ and $\dot{M}_{\rm g,pl}=2\times10^{-7}~M_{\rm J}~{\rm yr}^{-1}$. The sky-blue plots represent the Monte-Carlo simulations considering the parameter ranges listed in Tab. \ref{tab:montecarlo}. The open (colourful and grey) circles represent the cases where $Q_{\rm Toomre}<1$ at the outer regions of the discs. The green lines represent the observed value, $86\pm16~\mu{\rm Jy}$ \citep{ben21}.}
\label{fig:parameters}
\end{figure}

First, we investigate the total dust emission flux density from the CPD of PDS~70~c by changing the strength of turbulence as $\alpha=10^{-6}$, $10^{-5}$, $10^{-4}$, $10^{-3}$, and $10^{-2}$. We calculate 21 cases for each $\alpha$ by changing the value of $x$ from $0.001$ to $0.01$ at even intervals in a log scale. The colour plots in Fig. \ref{fig:parameters} shows that there is a negative correlation between the total dust emission and $\alpha$, which is explained as follows. The gas surface density is low when the turbulence is strong (and the gas accretion rate is fixed). Then, the Stokes number of dust is large, and the dust quickly drifts inwards, resulting in lower dust surface density. As a result, the total flux density of dust emission is small (see Section \ref{detailed} for the detailed explanation). The emission has a peak at $\alpha=10^{-5}$, because the orbital position of the transition from the Epsilon to Stokes regimes shifts outwards as $\alpha$ is small. The dust particles can grow faster in the Stokes regime than in the Epstein regime, which makes the drifting speed faster and the dust surface density lower, resulting in weaker dust emission. This result is roughly consistent with \citet{bae19} that the dust emission can be reproduced only when $\alpha\lesssim10^{-5}$, but the growth and radial drift of the dust are not considered in that work. The dependence on $x$ becomes inverse when $\alpha=10^{-6}$, which is also because the transition position shifts outwards as $x$ is large.

However, when $\alpha=10^{-6}$, the disc should be gravitationally unstable. We check the condition for the gravitational instability by calculating the Toomre Q parameter \citep[e.g.][]{too64},
\begin{equation}
Q_{\rm Toomre}\equiv\dfrac{c_{\rm s}\Omega_{\rm K}}{\pi G\Sigma_{\rm g}}.
\label{ToomreQ}
\end{equation}
When $\alpha=10^{-6}$, the gas surface density is very high and $Q_{\rm Toomre}$ is lower than unity at the outer region of the gas disc, which meets the condition for the gravitational instability (shown as open circles in Fig. \ref{fig:parameters}; see also Appendix \ref{GI}).

We then calculate Monte-Carlo simulations 1000 times by selecting random values of the parameters from the ranges listed in Tab. \ref{tab:montecarlo}. We also select the value of $\alpha$ and $x$ at random from the ranges of $10^{-6}\leq\alpha\leq10^{-2}$ and $0.0001\leq x\leq0.01$. The sky-blue and grey circles in Fig. \ref{fig:parameters} are the results of the Monte-Carlo simulations, which shows that the listed parameters do not affect the results so much. When about $\alpha<10^{-5}$, Toomre Q parameter is lower than unity at the outer region of the disc, suggesting the disc is gravitationally unstable (grey open circles).

For the Monte-Carlo simulations, we choose the parameter ranges listed in Tab. \ref{tab:montecarlo} because of the following reasons. The critical collision velocity for the fragmentation estimated by experiments is slower than that by numerical simulations \citep{blu00,wad13}. The estimated mass of the central star PDS~70 and its distance from the Earth have some ranges \citep{mul18,gaia16,gaia18}. The estimated inclination of the CPD is assumed to have the same inclination with the PPD. The inclination of the PPD depends on the position in the disc and the approaches of estimate used in previous works \citep{has15,kep19}. The estimated distance from PDS~70~c to the central star has a range \citep{haf19}. The estimated radius and surface temperature of the planet also have ranges because of the difference of models \citep{wan21}. The temperature of the PPD at the orbital position of PDS~70~c depends on the researches. We use the value of $32~{\rm K}$ estimated by \citet{law24}, but \citet{por23} estimates lower temperature, $16~{\rm K}$. Therefore, we change the temperature between the two estimated values.

\begin{table}
\caption{Parameter ranges}
\label{tab:montecarlo}
\centering
\begin{tabular}{lll}
\hline
Symbol & Ranges & Unit \\
\hline
$x$ & $10^{-4}-10^{-2}$ & - \\
$\alpha$ & $10^{-6}-10^{-2}$ & - \\
$v_{\rm ice}$ & $10-50$ & ${\rm m~s^{-1}}$ \\
$v_{\rm rock}$ & $1-5$ & ${\rm m~s^{-1}}$ \\
$q$ & $3.9-2.5$ & - \\
\hline
$M_{\rm star}$ & $0.74-0.78$ & $M_{\odot}$ \\
$d$ & $112.91-113.95$ & ${\rm pc}$ \\
$i$ & $49.7-51.7$ & ${\rm radian}$ \\
\hline
$a_{\rm pl}$ & $32.5-36.5$ & ${\rm au}$ \\
$R_{\rm pl}$ & $1.7-2.3$ & $R_{\rm J}$ \\
$T_{\rm pl,eff}$ & $1007-1113$ & ${\rm K}$ \\
$T_{\rm irr,PPD}$ & $16-32$ & ${\rm K}$ \\
\hline
\end{tabular}
\end{table}

\section{Discussion} \label{discussion}
\subsection{Constraints on planet properties}\label{constraints}
In Section \ref{planet-properties}, we show that the dust continuum emission from the CPD depends on the planet properties. Here, we obtain constraints on the properties by calculating the emission and comparing it with the observation data.

\subsubsection{Constraints on PDS~70~c} \label{constraints-PDS70c}
\begin{figure*}[htbp]
\centering
\includegraphics[width=0.55\linewidth]{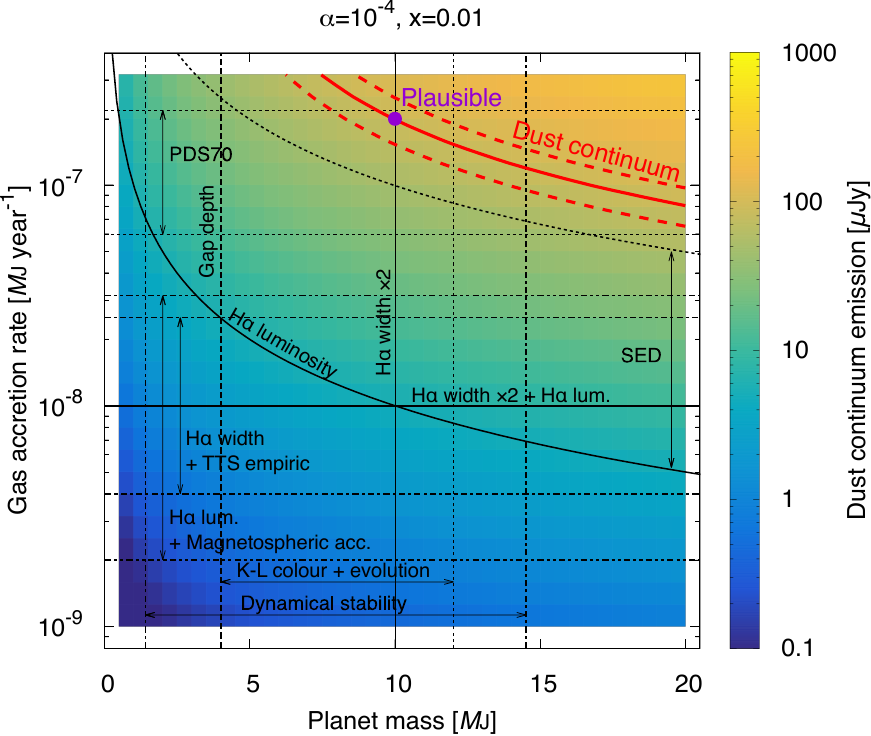}
\includegraphics[width=0.44\linewidth]{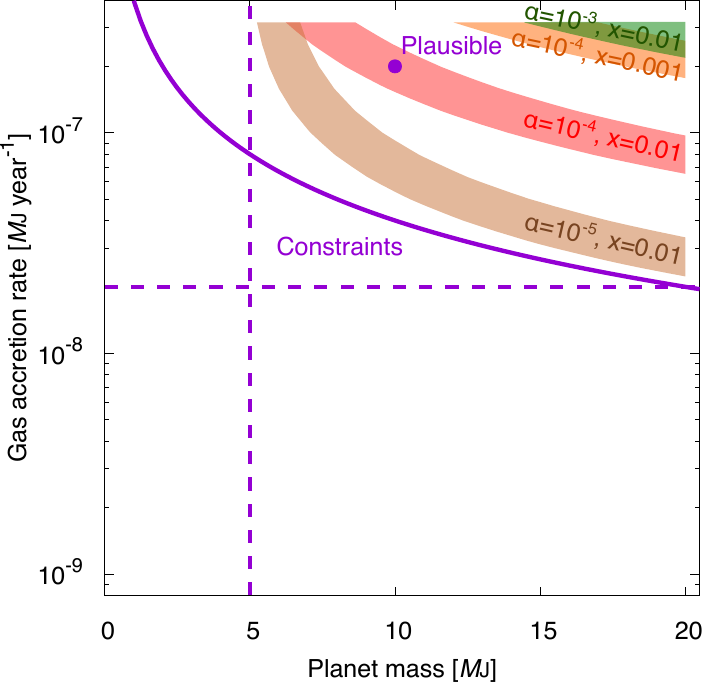}
\caption{Predicted flux density of dust emission from the CPD of PDS~70~c when the planet mass and gas accretion rate are $M_{\rm pl}=(0.5-20)~M_{\rm J}$ and $\dot{M}_{\rm g,pl}=(10^{-9}-10^{-6.5})~M_{\rm J}~{\rm yr}^{-1}$. The purple circles represent the plausible planet properties: $M_{\rm pl}=10~M_{\rm J}$ and $\dot{M}_{\rm g,pl}=2\times10^{-7}~M_{\rm J}~{\rm yr}^{-1}$. (Left) The strength of turbulence and the dust-to-gas mass ratio in the inflow are fixed as $\alpha=10^{-4}$ and $x=0.01$, respectively. The red solid and dashed curves represent the planet property range reproducing the observed value, $86\pm16~\mu{\rm Jy}$ \citep{ben21}. The black curves and lines represent the previous estimates of the properties listed in Tab. \ref{tab:estimates}. (Right) The red, brown, green, and orange shaded regions represent the planet property ranges reproducing the observed value with $(\alpha, x)=(10^{-4}, 0.01)$, $(10^{-5}, 0.01)$, $(10^{-3}, 0.01)$, and $(10^{-4}, 0.001)$, respectively. The solid purple curve is the obtained constraint on MMdot: $M_{\rm pl}\dot{M}_{\rm g,pl}\geq4\times10^{-7}~M_{\rm J}^{2}~{\rm yr}^{-1}$. The vertical and horizontal dashed purple lines represent the obtained constraints on the planet mass and the gas accretion rate, $M_{\rm pl}\geq5~M_{\rm J}$ and $\dot{M}_{\rm g,pl}\geq2\times10^{-8}~M_{\rm J}~{\rm yr}^{-1}$, respectively.}
\label{fig:constraints_PDS70c}
\end{figure*}

First, we investigate the constraints on the properties of PDS~70~c. The left panel of Fig. \ref{fig:constraints_PDS70c} shows the predicted flux density of dust emission from the CPD of PDS~70~c when the planet mass and the gas accretion rate are $M_{\rm pl}=(0.5-20)~M_{\rm J}$ and $\dot{M}_{\rm g,pl}=(10^{-9}-10^{-6.5})~M_{\rm J}~{\rm yr}^{-1}$. The colour contour shows that both of the planet mass and the gas accretion rate have positive correlation with the dust emission flux density, which is also shown in Section \ref{planet-properties}. The red curves represent the planet property range reproducing the observed flux density of dust emission $F_{\rm emit}=86\pm16~\mu{\rm Jy}$ \citep{ben21}. The black curves and lines represent the previous estimates of the planet mass and the gas accretion rate. Here, we assume that $\alpha=10^{-4}$, which is consistent with the theoretical prediction that magnetorotational instability (MRI) is not likely to occur in CPDs due to the short typical length scale of CPDs \citep{fuj14,tur14}. We also assume that the dust-to-gas mass ratio in the inflow is consistent with the stellar composition, $x=0.01$.

The red curves show that planet mass should be larger than about $5~M_{\rm J}$, which is consistent with most of the previous estimate. \citet{aoy19} estimates the mass of PDS~70~c as $10~M_{\rm J}$ from the combination of the 10\% full width and 50\% full width of the H$\alpha$ line (vertical solid line). \citet{haf19} also estimates the mass as $4-12~M_{\rm J}$ by comparing the K-L colour of the planet to evolutionary models of gas planets (between the vertical dotted and dashed lines. These previous estimated range is also in the range estimated by \citet{wan21} from the orbital dynamical stability with 95\% credible interval, $1.4-14.5~M_{\rm J}$ (between the vertical dot-dashed lines). \citet{por23} estimates the planet mass as $4~M_{\rm J}$ from the gap depth using the fitting formulas for multiple planet cases by \citet{duf15a}, which is smaller than our estimate (vertical dashed line). However, the gap structure also depends on the strength of turbulence (diffusion) and temperature around the gap, which are still poorly known.

The red curves also show that the dust emission can be reproduced when $\dot{M}_{\rm g,pl}\gtrsim5\times10^{-8}~M_{\rm J}~{\rm yr}^{-1}$. On the other hand, \citet{aoy19} estimates the gas accretion rate as $1\times10^{-8}~M_{\rm J}~{\rm yr}^{-1}$ from the combination of the 10\% full width, 50\% full width, and the luminosity of the H$\alpha$ line (the horizontal solid line). Also, only with the luminosity, MMdot is estimated as $1\times10^{-7}~M_{\rm J}^{2}~{\rm yr}^{-1}$ (the solid curve). \citet{wan21} also estimate the value of MMdot as $(1-10)\times10^{-7}~M_{\rm J}^{2}~{\rm yr}^{-1}$ by fitting a CPD model proposed by \citet{zhu15a} with the SED of infrared observations (between the dotted and solid curves). Although these previous estimates of $\dot{M}_{\rm g,pl}$ are lower than ours, the gas accretion rate and MMdot estimated by such observations can be larger in several orders of magnitude if the extinction by dust is considered, which reduces the apparent luminosity of the object \citep{has20,mar22}\footnote{We note that \citet{has20} estimates the lower limits of the degree of the extinction from the flux ratio of H$\alpha$ and H$\beta$ (non-detection), but the used data includes instrumental uncertainties which may cause overestimates. Therefore, we do not use their estimated value in our discussion.}. \citet{tha19} estimates the gas accretion rate as $1\times10^{-8.1\pm0.6}~M_{\rm J}~{\rm yr}^{-1}$ with the assumption of $M_{\rm pl}=6~M_{\rm J}$ by applying a magnetospheric accretion model of T Tauri stars (TTSs) by \citet{muz01} to the H$\alpha$ luminosity of PDS~70~c (between the dot-dashed horizontal lines). On the other hand, by fitting the magnetospheric accretion model to the H$\alpha$ line profiles of PDS~70 (central star), \citet{tha20} estimates the gas accretion rate of the star as $(1.4\pm0.8)\times10^{-7}~M_{\rm J}~{\rm yr}^{-1}$ (between the dotted horizontal lines). This should be the upper limit of the gas accretion rate of PDS~70~c, because hydrodynamical simulations show that the gas accretion rate onto gas planets can be about 90\% of that from the outer part of the PPDs \citep{lub06}. We note that our estimate is outside the previous estimate by \citet{haf19}: $\dot{M}_{\rm g}=1\times10^{-8\pm0.4}~M_{\rm J}~{\rm yr}^{-1}$ (between the dot-dashed horizontal lines). This estimate is obtained by the empirical relation between the gas accretion rate and the H$\alpha$ 10\% width, which should not be affected by the extinction. However, the used empirical relation was obtained from observations of T Tauri stars and brown dwarfs \citep{nat04}, which could underestimate the gas accretion rate \citep{tha19,aoy21}. Also, the flow pattern around a planet surface is still controversial, which is a sensitive factor for the H$\alpha$ emission \citep{tak21,mar23}.

Considering the above comparisons, we estimate the plausible planet mass and the gas accretion rate of PDS~70~c as $10~M_{\rm J}$ and $2\times10^{-7}~M_{\rm J}~{\rm yr}^{-1}$ (purple circles). In this case, MMdot is $M_{\rm pl}\dot{M}_{\rm g,pl}=2\times10^{-6}~M_{\rm J}^{2}{\rm yr}^{-1}$, which is consistent with the estimate from H$\alpha$ luminosity by \citet{aoy19}, $M_{\rm pl}\dot{M}_{\rm g,pl}=1\times10^{-7}~M_{\rm J}^{2}{\rm yr}^{-1}$, with the extinction factor of $A_{\rm H\alpha}'=\log_{10}20=1.3$. Also, the total mass of the dust inside the CPD is $M_{\rm d}=0.014~M_{\oplus}$ when $\alpha=10^{-4}$ and $x=0.01$, which is inside the ranges of the dust mass estimated by \citet{ben21}, $M_{\rm d}\sim0.007-0.031~M_{\oplus}$.

The right panel of Fig. \ref{fig:constraints_PDS70c} shows the planet property ranges reproducing the observed flux density of dust emission with the various strength of turbulence in the CPDs and the dust-to-gas mass ratio of the gas inflow; $(\alpha, x)=(10^{-4}, 0.01)$ (red), $(10^{-5}, 0.01)$ (brown), $(10^{-3}, 0.01)$ (green), and $(10^{-4}, 0.001)$ (orange). 

In Section \ref{planet-properties}, we show that the flux density of dust emission has positive correlations with the planet mass, the gas accretion rate, and their product MMdot. We also show that the flux density takes the highest value when $\alpha=10^{-5}$ in Section \ref{properties}. In that section, we also show that the flux density has a positive correlation with the dust-to-gas mass ratio in the gas inflow as well (when $\alpha\geq10^{-5}$). The mass ratio must be lower than the stellar composition, $x=0.01$, because the gas pressure bump at the outer edge of the gap halts the radial drift of the dust in the PPD \citep{zhu12,kan18,bae19,hom20,kar23}\footnote{\citet{szu22} finds that the meridional circulation can bring dust to CPDs by carrying out 3D dust+gas radiative hydrodynamic simulation, but the amount of supply depends on the simulation settings.}. Therefore, we can regard the planet properties reproducing the observed flux density of dust emission with $(\alpha, x)=(10^{-5}, 0.01)$ as their lower limits. We then obtain constraints on the MMdot of PDS~70~c: $M_{\rm pl}\dot{M}_{\rm g,pl}\geq4\times10^{-7}~M_{\rm J}^{2}~{\rm yr}^{-1}$ (solid purple curve). We also obtain the constraints on the planet mass and the gas accretion rate: $M_{\rm pl}\geq5~M_{\rm J}$ and $\dot{M}_{\rm g,pl}\geq2\times10^{-8}~M_{\rm J}~{\rm yr}^{-1}$.

\subsubsection{Constraints on PDS~70~b}\label{constraints-PDS70b}
\begin{figure*}[htbp]
\centering
\includegraphics[width=0.55\linewidth]{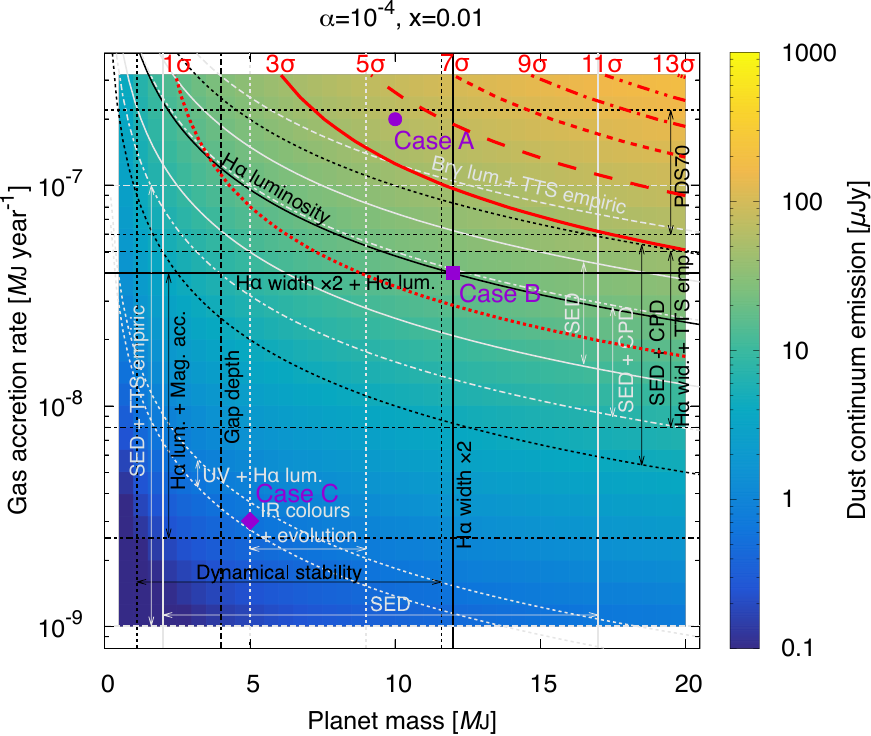}
\includegraphics[width=0.44\linewidth]{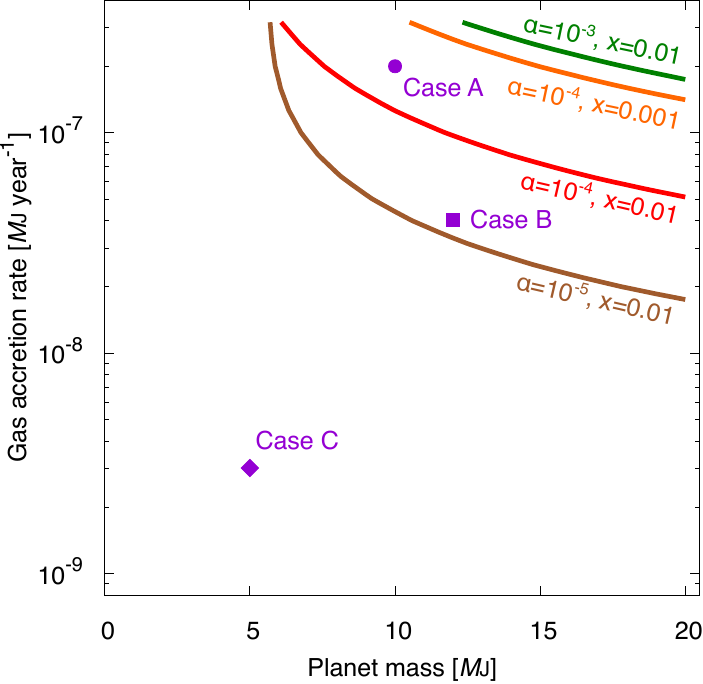}
\caption{Same as Fig. \ref{fig:constraints_PDS70c} but for PDS~70~b. The purple circles, squares, and diamonds respectively represent the planet properties of the three cases; Case A ($M_{\rm pl}=10~M_{\rm J}$ and $\dot{M}_{\rm g,pl}=2\times10^{-7}~M_{\rm J}~{\rm yr}^{-1}$; same with the plausible case of PDS~70~c), Case B ($M_{\rm pl}=12~M_{\rm J}$ and $\dot{M}_{\rm g,pl}=4\times10^{-8}~M_{\rm J}~{\rm yr}^{-1}$), and Case C ($M_{\rm pl}=5~M_{\rm J}$ and $\dot{M}_{\rm g,pl}=3\times10^{-9}~M_{\rm J}~{\rm yr}^{-1}$). (Left) The red curves represent the signal to noise ratios ($1\sigma=15.7~\mu{\rm Jy}$) if the dust emission were detected in the observation by \citet{ben21}. The black curves and lines represent the previous estimates of the properties listed in Tab. \ref{tab:estimates}. The light grey curves and lines also represent the previous estimates, but they are from different works from those of PDS~70~c. (Right) The red, brown, green, and orange curves represent the $3\sigma=47.1~\mu{\rm Jy}$ noise levels with $(\alpha, x)=(10^{-4}, 0.01)$, $(10^{-5}, 0.01)$, $(10^{-3}, 0.01)$, and $(10^{-4}, 0.001)$, respectively.}
\label{fig:constraints_PDS70b}
\end{figure*}

Next, we consider the constraints on the properties of PDS~70~b. There have been no detection of dust continuum emission from the CPD of PDS~70~b, but the non-detection can provide us loose upper limits of the planet properties. The left panel of Fig. \ref{fig:constraints_PDS70b} shows the predictions of the dust continuum emission with the same ranges of the planet properties as those of Fig. \ref{fig:constraints_PDS70c}. The panel shows that both of the planet mass and the gas accretion rate have positive correlation with the dust emission flux density as well as those of PDS~70~c shown in Figs. \ref{fig:planet-properties} and \ref{fig:constraints_PDS70c}. The red curves represent the noise levels of the observation of PDS~70~b by \citet{ben21} in every $1\sigma=15.7~\mu{\rm Jy}$. The non-detection requires that the planet mass and the gas accretion rate are lower than the red curves.

The black curves and lines represent the previous estimates of the properties of PDS~70~b listed in Tab. \ref{tab:estimates}. The light grey ones also represent the previous estimates of the planet, but they does not estimate planet c in their works. \citet{aoy19} estimates the mass as $12~M_{\rm J}$ from 10\% full width and 50\% full width of the H$\alpha$ line (black vertical solid line). \citet{mul18} also estimates the mass as $2-17~M_{\rm J}$ from the SED of infrared (between the black vertical doted lines), and that range covers the estimate by \citet{aoy19}. The mass range estimated by \citet{wan21} from the orbital dynamical stability with 95\% credible interval is $1.1-11.6~M_{\rm J}$ (between the black vertical dot-dashed lines), but relatively small mass in the range is more likely. \citet{kep18} also estimates the mass as $5-9~M_{\rm J}$ considering a formation model (between the grey dashed lines). Moreover, \citet{por23} estimates the mass as $4~M_{\rm J}$ from the gap depth (black dashed line).

\citet{chr19a} estimates the upper limit of the MMdot from Br$\gamma$ line luminosity and its empirical relation to the gas accretion rate for TTSs as $<1.26\times10^{-6}~M_{\rm J}^{2}~{\rm yr}^{-1}$ when the planet radius is $2.0~R_{\rm J}$ (grey dashed curve). \citet{chr19b} also estimates the MMdot as $10^{-6.8}-10^{-6.3}~M_{\rm J}^{2}~{\rm yr}^{-1}$ from the SED of infrared with a CPD model produced by \citet{eis15} (between the grey dot-dashed curves). \citet{aoy19} estimates the gas accretion rate as $4\times10^{-8}~M_{\rm J}~{\rm yr}^{-1}$ from 10\% full width, 50\% full width, and the luminosity of the H$\alpha$ line (horizontal black solid line). Only with the luminosity, \citet{aoy19} estimates the MMdot as $4.8\times10^{-7}~M_{\rm J}^{2}~{\rm yr}^{-1}$ (black solid curve). \citet{wan21} estimates the value of MMdot as $(1-10)\times10^{-7}~M_{\rm J}^{2}~{\rm yr}^{-1}$ by fitting the CPD model with the SED of infrared (between the black dotted curves). \citet{sto20} estimates the MMdot as $(2.5-7.5)\times10^{-7}~M_{\rm J}^{2}~{\rm yr}^{-1}$ with the SED of the infrared emission (between the grey dot-dashed curves). \citet{wag18} estimates the gas accretion rate as $1\times10^{-8\pm1}~M_{\rm J}~{\rm yr}^{-1}$, but it uses an empirical relation of TTSs \citep{rig12}, so that the estimate could be an underestimate \citep{tha19} (between the black horizontal dot-dashed lines)\footnote{To be accurate, the property directly estimated by the luminosity of H$\alpha$ line is not the gas accretion rate but MMdot. However, the variation of the gas accretion rate is much larger than that of the planet mass, making this estimate acceptable.}. \citet{tha19} estimates the gas accretion rate as $1\times10^{-8.1\pm0.6}~M_{\rm J}~{\rm yr}^{-1}$ with the assumption of $M_{\rm pl}=6~M_{\rm J}$ by applying the magnetospheric accretion model for TTSs to the H$\alpha$ luminosity of PDS~70~b (between the horizontal black dot-dot-dashed lines). An estimate by \citet{haf19} using the H$\alpha$ 10\% width and the empirical relation of TTS and blown dwarfs is $2\times10^{-8\pm0.4}~M_{\rm J}~{\rm yr}^{-1}$ (between the black dot-dashed curves). \citet{zho21} estimates lower value of MMdot: $(1.6\pm0.23)\times10^{-8}~M_{\rm J}^{2}~{\rm yr}^{-1}$ by the luminosity of ultraviolet (UV) and H$\alpha$ line obtained by the Hubble Space Telescope observation (between the light grey solid curves). The gas accretion rate of the central star, $(1.4\pm0.8)\times10^{-7}~M_{\rm J}~{\rm yr}^{-1}$ \citep{tha20}, also provides us the upper limit of the gas accretion rate of PDS~70~b (between the horizontal black dotted lines).

Here, we consider the following three plausible cases of PDS~70~b. First, we consider a case where the planet mass and the gas accretion rate are the same with those of the `plausible case' of PDS~70~c (Case A: purple circles). This is consistent with the estimate from H$\alpha$ luminosity by \citet{aoy19} with the extinction factor of $A_{\rm H\alpha}'=\log_{10}(20.0/4.8)=0.62$. In this case, the left panel of Fig. \ref{fig:constraints_PDS70b} shows that the dust emission should have been detected with $>3\sigma$ when $\alpha=10^{-4}$ and $x=0.01$, which is inconsistent with the non-detection \citep{ben21}. The right panel shows that the dust emission should have also been detected with $>3\sigma$ when $\alpha=10^{-5}$ and $x=0.01$. On the other hand, the panel shows that the predicted dust emission is lower than $3\sigma$ when the turbulence is strong ($\alpha=10^{-3}$) or the dust-to-gas mass ratio in the gas inflow is low ($x=0.001$), which is consistent with the non-detection.

Second, if the estimate by \citet{aoy19} is true and there is no dust extinction, the planet mass and the gas accretion rate of PDS~70~b are $12~M_{\rm J}$ and $4\times10^{-8}~M_{\rm J}~{\rm yr}^{-1}$, respectively (Case B: purple squares). In this case, the left panel shows that the dust emission should have been detected only with $1-3\sigma$ when $\alpha=10^{-4}$ and $x=0.01$. The right panel shows that the emission should have been $3\sigma$ detection when $\alpha=10^{-5}$ and $x=0.01$, which is inconsistent with the observation. On the other hand, the predicted dust emission is lower than the $3\sigma$ when $\alpha\leq10^{-3}$ or $x\leq0.001$.

Finally, we consider a case where the planet mass and the gas accretion rate are $5~M_{\rm J}$ and $4\times10^{-8}~M_{\rm J}~{\rm yr}^{-1}$, respectively (Case C; purple diamonds). Numerical orbital simulations show that the two planets can have such close orbits, because they are in 2:1 mean motion resonance (with relatively high eccentricity of planet b, $e\sim0.1$), and their orbits are more stable when the planet b has smaller mass than planet c, $M_{\rm pl,b}\lesssim5~M_{\rm J}$, compared to when both of the planets have large mass \citep{bae19,wan21}. Such small mass of planet b is also suggested by planet evolution models and its gap depth \citep{kep18,por23}. The estimate from UV and H$\alpha$ luminosity also gives much smaller MMdot of the planet than the other previous estimates \citep{zho21}. In Case C, the left panel shows that the the predicted dust emission is much lower than the $1\sigma$ noise level when $\alpha=10^{-4}$ and $x=0.01$. The right panel also shows that the emission should be much lower than the $3\sigma$ with any set of $\alpha$ and $x$.

\subsection{Possible scenarios for PDS~70~b and c} \label{scenario}
There have been two embedded planets discovered in PDS~70 system, and both of the planets are accreting gas. However, only the dust continuum emission from the CPD of PDS~70~c, the outer planet, has been detected in the previous observations, and its reason is still unknown. From the constraints obtained in Section \ref{constraints}, we propose two possible scenarios to explain the reason.

The first possibility is that planet c has larger planet mass, gas accretion rate, or both than planet b. In Section \ref{planet-properties}, we found that the flux density of dust emission is about proportional to the planet mass, the gas accretion rate, and their product, MMdot. As a result, we obtained their lower limits from the detected value of the dust emission in Section \ref{constraints-PDS70c}; $M_{\rm pl}\dot{M}_{\rm g,pl}\geq4\times10^{-7}~M_{\rm J}^{2}~{\rm yr}^{-1}$, $M_{\rm pl}\geq5~M_{\rm J}$, and $\dot{M}_{\rm g,pl}\geq2\times10^{-8}~M_{\rm J}~{\rm yr}^{-1}$. On the other hand, we showed that the dust emission of planet b is lower than the detection limit if the planet mass, the gas accretion rate, or both are low enough in Section \ref{constraints-PDS70b}. For example, the predicted flux density of dust emission is lower than the $3\sigma=47.1~\mu{\rm Jy}$ of the observation in Case B ($M_{\rm pl}=12~M_{\rm J}$ and $\dot{M}_{\rm g,pl}=4\times10^{-8}~M_{\rm J}~{\rm yr}^{-1}$) and Case C ($M_{\rm pl}=5~M_{\rm J}$ and $\dot{M}_{\rm g,pl}=3\times10^{-9}~M_{\rm J}~{\rm yr}^{-1}$), and these values of the properties are supported by previous researches of planet b using other methods. The properties of Case B are consistent with the observed linewidth and luminosity of H$\alpha$ emission \citep{aoy19}. The properties of Case C are consistent with the orbital stability \citep{bae19,wan21}, planet evolution \citep{kep19}, gap depth \citep{por23}, and another UV and H$\alpha$ luminosity observation \citep{zho21}. However, this scenario cannot directly explain why planet b has stronger H$\alpha$ luminosity than planet c \citep{aoy21}.

The other possibility is that planet c has a stronger turbulence in the CPD, higher dust-to-gas mass ratio in the gas inflow onto the CPD, or both. As shown in Section \ref{properties}, there is a negative correlation between the strength of turbulence and the dust emission flux density, and a positive correlation between the dust-to-gas mass ratio and the dust emission flux density. Figure \ref{fig:constraints_PDS70c} shows that the observed value of PDS~70~c can be reproduced only when $\alpha=10^{-4}$ and $x=0.01$ in the plausible planet properties case. On the other hand, Fig. \ref{fig:constraints_PDS70b} shows that the predicted flux density of PDS~70~b can be lower than the detection limit ($3\sigma$) if $\alpha=10^{-3}$ or $x=0.001$ in Case A, where planet b has the same planet properties with planet c. The turbulence in CPDs is unlikely strong because the condition for MRI is not easy to be achieved \citep{fuj14,tur14}. However, the strength of turbulence should depend on planet properties in reality, and the gas angular momentum transfer can be driven by other mechanisms such as spiral arms and magnetic disc wind \citep{zhu16,shi23}. Therefore, if such mechanisms work better in the CPD of planet b, the non-detection and detection of the planets b and c can be explained. Also, it is understandable that the inflow of planet c has higher dust-to-gas mass ratio than planet b, because the orbit of planet b is farther than that of planet c from the observed outer dust ring. By simple thinking, it is more difficult to supply the dust piled-up at the edge of the gap to the vicinity of planet b than to the planet c by any mechanisms such as the dust diffusion, the inwards gas flow, or the meridional circulation \citep{zhu12,kan18,hom20,szu22,kar23}. This picture is actually consistent with an interpretation of JWST/NIRCam images of PDS~70 system obtained recently \citep{chr24}.
The filtering of the radially drifting dust by the planet c may also reduce the supply of dust to the planet b. This relatively dust-rich environment of PDS~70~c can also qualitatively explain why the observed H$\alpha$ luminosity of planet c is lower than that of b, because an enough amount of dust can shade the emission from the planet and CPD, which is known as the dust extinction effect \citep{aoy19,wan21}.

\section{Conclusions} \label{conclution}
A forming planet embedded in a protoplanetary disc (PPD) accretes gas and forms a small gas disc called circumplanetary disc (CPD) around the planet. An extrasolar system PDS~70 has a PPD and two gas accreting planets, PDS~70~b and c. The dust continuum emission from the CPD of PDS~70~c has been detected by ALMA Band 7 ($\lambda=855~\mu{\rm m}$) but not from PDS~70~b \citep{ben21}. In this work, we obtained constraints on properties of the two planets by comparing the predicted dust emission by our model with the detected and non-detected dust emission from the CPDs. We modelled a 1D viscous accretion disc with inflow as a CPD formed around a gas accreting planet. We then modelled the evolution of dust inside the CPD, where the dust is supplied with the gas inflow, and the thermal emission from the dust inside the CPD, while previous works had not considered the dust evolution.

First, we investigated the dependence of the flux density of dust emission from CPDs on planet and CPD properties. We found that the flux density of the dust emission, $F_{\rm emit}$, depends on three fundamental properties of forming gas planets: the planet mass ($M_{\rm pl}$), the gas accretion rate ($\dot{M}_{\rm g, pl}$), and their product called MMdot ($M_{\rm pl}\dot{M}_{\rm g, pl}$). We showed that the flux density is almost proportional to the planet mass and the gas accretion rate; $F_{\rm emit}\propto M_{\rm pl}$ and $F_{\rm emit}\propto\dot{M}_{\rm g, pl}$. Therefore, the flux density of dust emission is also about proportional to MMdot ($F_{\rm emit}\propto M_{\rm pl}\dot{M}_{\rm g, pl}$), which suggests that MMdot is one of the most essential parameters as well as $M_{\rm pl}$ and $\dot{M}_{\rm g, pl}$ in the context of the dust emission. We also found that the strength of turbulence in CPDs ($\alpha$) and the dust-to-gas mass ratio in the inflow to CPDs ($x$) are important as well. The dust emission flux density has a peak at $\alpha=10^{-5}$, and the correlations between $F_{\rm emit}$ and $\alpha$ is negative and between $F_{\rm emit}$ and $x$ is positive when $\alpha\geq10^{-5}$. The correlations are opposite when $\alpha<10^{-5}$ due to the shift of the orbital position where the dust goes from the Epstein to Stokes regimes.

With this dependence on the planet properties, we then obtained the constraints on PDS~70~b and c by investigating the conditions for reproducing the observed value of the dust emission from PDS~70~c, $86\pm16~\mu{\rm m}$, and the non-detection of planet b. First, we constrained the properties of PDS~70~c as $M_{\rm pl}\dot{M}_{\rm g,pl}\geq4\times10^{-7}~M_{\rm J}^{2}~{\rm yr}^{-1}$, corresponding to $M_{\rm pl}\geq5~M_{\rm J}$ and $\dot{M}_{\rm g,pl}\geq2\times10^{-8}~M_{\rm J}~{\rm yr}^{-1}$. This is the first case to succeed in obtaining the constraints on the properties of gas accreting planets from the flux density of dust emission from the CPDs. We estimated the plausible case of the properties as $M_{\rm pl}=10~M_{\rm J}$ and $\dot{M}_{\rm g,pl}=2\times10^{-7}~M_{\rm J}~{\rm yr}^{-1}$ by comparing our results with the previous estimates using other methods or observations. We also found there are two possibilities for PDS~70~b by considering the conditions for reproducing the non-detection ($F_{\rm emit}<3\sigma=47.1~\mu{\rm Jy}$). The first possibility is that planet b has smaller mass, lower gas accretion rate, or both than planet c. The other possibility is that PDS~70~b has stronger CPD turbulence, lower dust-to-gas mass ratio in the inflow, or both. The scenario that PDS~70~c has a larger amount of dust supply than PDS~70~b is consistent with the fact that planet c is closer to the outer dust ring than planet b, and the relatively dust-rich environment of planet c can quantitatively explain why the luminosity of H$\alpha$ emission of the planet is lower than that of planet b by the dust extinction effect.

\begin{acknowledgements}
We thank the anonymous referee for the very valuable comments, which improved our manuscript a lot. We thank Yann Alibert for very useful and constructive discussion thorough the whole this research. We also thank Jun Hashimoto and Yuhiko Aoyama for very important comments from the observation aspect. We appreciate Takahiro Ueda giving us useful advice on the development of the dust emission model. This work has been carried out within the framework of the NCCR PlanetS supported by the Swiss National Science Foundation under grants 51NF40\_182901 and 51NF40\_205606. C.M. acknowledges the funding from the Swiss National Science Foundation under grant 200021\_204847 `Planets In Time’. This work was supported by JSPS KAKENHI Grant Number JP22H01274.
\end{acknowledgements}

\bibliographystyle{aa}
\bibliography{cpd_emission}

\begin{thebibliography}{91}
\expandafter\ifx\csname natexlab\endcsname\relax\def\natexlab#1{#1}\fi

\bibitem[{Adachi {et~al.}(1976)Adachi, Hayashi, \& Nakazawa}]{ada76}
Adachi, I., Hayashi, C., \& Nakazawa, K. 1976, Progress of Theoretical Physics,
  56, 1756

\bibitem[{Andrews {et~al.}(2021)Andrews, Elder, Zhang, Huang, Benisty,
  Kurtovic, Wilner, Zhu, Carpenter, P{\'e}rez, {et~al.}}]{and21}
Andrews, S.~M., Elder, W., Zhang, S., {et~al.} 2021, The Astrophysical Journal,
  916, 51

\bibitem[{Aoyama \& Ikoma(2019)}]{aoy19}
Aoyama, Y. \& Ikoma, M. 2019, The Astrophysical Journal Letters, 885, L29

\bibitem[{Aoyama {et~al.}(2021)Aoyama, Marleau, Ikoma, \& Mordasini}]{aoy21}
Aoyama, Y., Marleau, G.-D., Ikoma, M., \& Mordasini, C. 2021, The Astrophysical
  Journal Letters, 917, L30

\bibitem[{Bae {et~al.}(2022)Bae, Teague, Andrews, Benisty, Facchini,
  Galloway-Sprietsma, Loomis, Aikawa, Alarc{\'o}n, Bergin, {et~al.}}]{bae22}
Bae, J., Teague, R., Andrews, S.~M., {et~al.} 2022, The Astrophysical Journal
  Letters, 934, L20

\bibitem[{Bae {et~al.}(2019)Bae, Zhu, Baruteau, Benisty, Dullemond, Facchini,
  Isella, Keppler, P{\'e}rez, \& Teague}]{bae19}
Bae, J., Zhu, Z., Baruteau, C., {et~al.} 2019, The Astrophysical Journal
  Letters, 884, L41

\bibitem[{Balsalobre-Ruza {et~al.}(2023)Balsalobre-Ruza, de~Gregorio-Monsalvo,
  Lillo-Box, Hu{\'e}lamo, Ribas, Benisty, Bae, Facchini, \& Teague}]{bal23}
Balsalobre-Ruza, O., de~Gregorio-Monsalvo, I., Lillo-Box, J., {et~al.} 2023,
  Astronomy \& Astrophysics

\bibitem[{{Bauer} {et~al.}(1997){Bauer}, {Finocchi}, {Duschl}, {Gail}, \&
  {Schloeder}}]{bau97}
{Bauer}, I., {Finocchi}, F., {Duschl}, W.~J., {Gail}, H.~P., \& {Schloeder},
  J.~P. 1997, \aap, 317, 273

\bibitem[{Bell \& Lin(1994)}]{bel94}
Bell, K. \& Lin, D. 1994, Astrophysical Journal, Part 1 (ISSN 0004-637X), vol.
  427, no. 2, p. 987-1004, 427, 987

\bibitem[{Benisty {et~al.}(2021)Benisty, Bae, Facchini, Keppler, Teague,
  Isella, Kurtovic, P{\'e}rez, Sierra, Andrews, {et~al.}}]{ben21}
Benisty, M., Bae, J., Facchini, S., {et~al.} 2021, The Astrophysical Journal
  Letters, 916, L2

\bibitem[{Birnstiel {et~al.}(2018)Birnstiel, Dullemond, Zhu, Andrews, Bai,
  Wilner, Carpenter, Huang, Isella, Benisty, {et~al.}}]{bir18}
Birnstiel, T., Dullemond, C.~P., Zhu, Z., {et~al.} 2018, The Astrophysical
  Journal Letters, 869, L45

\bibitem[{Blum \& Wurm(2000)}]{blu00}
Blum, J. \& Wurm, G. 2000, Icarus, 143, 138

\bibitem[{Canup \& Ward(2002)}]{can02}
Canup, R.~M. \& Ward, W.~R. 2002, The Astronomical Journal, 124, 3404

\bibitem[{Canup \& Ward(2006)}]{can06}
Canup, R.~M. \& Ward, W.~R. 2006, Nature, 441, 834

\bibitem[{Casassus \& C{\'a}rcamo(2022)}]{cas22}
Casassus, S. \& C{\'a}rcamo, M. 2022, Monthly Notices of the Royal Astronomical
  Society, 513, 5790

\bibitem[{Chiang \& Goldreich(1997)}]{chi97}
Chiang, E. \& Goldreich, P. 1997, The Astrophysical Journal, 490, 368

\bibitem[{Chiang \& Youdin(2010)}]{chi10}
Chiang, E. \& Youdin, A. 2010, Annual Review of Earth and Planetary Sciences,
  38, 493

\bibitem[{Christensen {et~al.}(2009)Christensen, Holzwarth, \& Reiners}]{chr09}
Christensen, U.~R., Holzwarth, V., \& Reiners, A. 2009, Nature, 457, 167

\bibitem[{Christiaens {et~al.}(2019{\natexlab{a}})Christiaens, Cantalloube,
  Casassus, Price, Absil, Pinte, Girard, \& Montesinos}]{chr19b}
Christiaens, V., Cantalloube, F., Casassus, S., {et~al.} 2019{\natexlab{a}},
  The Astrophysical Journal Letters, 877, L33

\bibitem[{Christiaens {et~al.}(2019{\natexlab{b}})Christiaens, Casassus, Absil,
  Cantalloube, Gomez~Gonzalez, Girard, Ram{\'\i}rez, Pairet, Salinas, Price,
  {et~al.}}]{chr19a}
Christiaens, V., Casassus, S., Absil, O., {et~al.} 2019{\natexlab{b}}, Monthly
  Notices of the Royal Astronomical Society, 486, 5819

\bibitem[{Christiaens {et~al.}(2024)Christiaens, Samland, Henning,
  Portilla-Revelo, Perotti, Matthews, Absil, Decin, Kamp, Boccaletti,
  {et~al.}}]{chr24}
Christiaens, V., Samland, M., Henning, T., {et~al.} 2024, arXiv preprint
  arXiv:2403.04855

\bibitem[{D'Angelo \& Spruit(2010)}]{dan10}
D'Angelo, C.~R. \& Spruit, H.~C. 2010, Monthly Notices of the Royal
  Astronomical Society, 406, 1208

\bibitem[{Duffell \& Dong(2015)}]{duf15a}
Duffell, P.~C. \& Dong, R. 2015, The Astrophysical Journal, 802, 42

\bibitem[{Eisner(2015)}]{eis15}
Eisner, J. 2015, The Astrophysical Journal Letters, 803, L4

\bibitem[{Emsenhuber {et~al.}(2021)Emsenhuber, Mordasini, Burn, Alibert, Benz,
  \& Asphaug}]{ems21a}
Emsenhuber, A., Mordasini, C., Burn, R., {et~al.} 2021, Astronomy \&
  Astrophysics, 656, A69

\bibitem[{Freedman {et~al.}(2014)Freedman, Lustig-Yaeger, Fortney, Lupu,
  Marley, \& Lodders}]{fre14}
Freedman, R.~S., Lustig-Yaeger, J., Fortney, J.~J., {et~al.} 2014, The
  Astrophysical Journal Supplement Series, 214, 25

\bibitem[{Fujii {et~al.}(2014)Fujii, Okuzumi, Tanigawa, \& ichiro
  Inutsuka}]{fuj14}
Fujii, Y.~I., Okuzumi, S., Tanigawa, T., \& ichiro Inutsuka, S. 2014, The
  Astrophysical Journal, 785, 101

\bibitem[{{Gaia Collaboration} {et~al.}(2018){Gaia Collaboration}, {Brown},
  {Vallenari}, {Prusti}, {de Bruijne}, {Babusiaux}, {Bailer-Jones}, {Biermann},
  {Evans}, {Eyer}, {Jansen}, {Jordi}, {Klioner}, {Lammers}, {Lindegren},
  {Luri}, {Mignard}, {Panem}, {Pourbaix}, {Randich}, {Sartoretti}, {Siddiqui},
  {Soubiran}, {van Leeuwen}, {Walton}, {Arenou}, {Bastian}, {Cropper},
  {Drimmel}, {Katz}, {Lattanzi}, {Bakker}, {Cacciari}, {Casta{\~n}eda},
  {Chaoul}, {Cheek}, {De Angeli}, {Fabricius}, {Guerra}, {Holl}, {Masana},
  {Messineo}, {Mowlavi}, {Nienartowicz}, {Panuzzo}, {Portell}, {Riello},
  {Seabroke}, {Tanga}, {Th{\'e}venin}, {Gracia-Abril}, {Comoretto},
  {Garcia-Reinaldos}, {Teyssier}, {Altmann}, {Andrae}, {Audard},
  {Bellas-Velidis}, {Benson}, {Berthier}, {Blomme}, {Burgess}, {Busso},
  {Carry}, {Cellino}, {Clementini}, {Clotet}, {Creevey}, {Davidson}, {De
  Ridder}, {Delchambre}, {Dell'Oro}, {Ducourant},
  {Fern{\'a}ndez-Hern{\'a}ndez}, {Fouesneau}, {Fr{\'e}mat}, {Galluccio},
  {Garc{\'\i}a-Torres}, {Gonz{\'a}lez-N{\'u}{\~n}ez}, {Gonz{\'a}lez-Vidal},
  {Gosset}, {Guy}, {Halbwachs}, {Hambly}, {Harrison}, {Hern{\'a}ndez},
  {Hestroffer}, {Hodgkin}, {Hutton}, {Jasniewicz}, {Jean-Antoine-Piccolo},
  {Jordan}, {Korn}, {Krone-Martins}, {Lanzafame}, {Lebzelter}, {L{\"o}ffler},
  {Manteiga}, {Marrese}, {Mart{\'\i}n-Fleitas}, {Moitinho}, {Mora}, {Muinonen},
  {Osinde}, {Pancino}, {Pauwels}, {Petit}, {Recio-Blanco}, {Richards},
  {Rimoldini}, {Robin}, {Sarro}, {Siopis}, {Smith}, {Sozzetti}, {S{\"u}veges},
  {Torra}, {van Reeven}, {Abbas}, {Abreu Aramburu}, {Accart}, {Aerts},
  {Altavilla}, {{\'A}lvarez}, {Alvarez}, {Alves}, {Anderson}, {Andrei},
  {Anglada Varela}, {Antiche}, {Antoja}, {Arcay}, {Astraatmadja}, {Bach},
  {Baker}, {Balaguer-N{\'u}{\~n}ez}, {Balm}, {Barache}, {Barata}, {Barbato},
  {Barblan}, {Barklem}, {Barrado}, {Barros}, {Barstow}, {Bartholom{\'e}
  Mu{\~n}oz}, {Bassilana}, {Becciani}, {Bellazzini}, {Berihuete}, {Bertone},
  {Bianchi}, {Bienaym{\'e}}, {Blanco-Cuaresma}, {Boch}, {Boeche}, {Bombrun},
  {Borrachero}, {Bossini}, {Bouquillon}, {Bourda}, {Bragaglia}, {Bramante},
  {Breddels}, {Bressan}, {Brouillet}, {Br{\"u}semeister}, {Brugaletta},
  {Bucciarelli}, {Burlacu}, {Busonero}, {Butkevich}, {Buzzi}, {Caffau},
  {Cancelliere}, {Cannizzaro}, {Cantat-Gaudin}, {Carballo}, {Carlucci},
  {Carrasco}, {Casamiquela}, {Castellani}, {Castro-Ginard}, {Charlot},
  {Chemin}, {Chiavassa}, {Cocozza}, {Costigan}, {Cowell}, {Crifo}, {Crosta},
  {Crowley}, {Cuypers}, {Dafonte}, {Damerdji}, {Dapergolas}, {David}, {David},
  {de Laverny}, {De Luise}, {De March}, {de Martino}, {de Souza}, {de Torres},
  {Debosscher}, {del Pozo}, {Delbo}, {Delgado}, {Delgado}, {Di Matteo},
  {Diakite}, {Diener}, {Distefano}, {Dolding}, {Drazinos}, {Dur{\'a}n},
  {Edvardsson}, {Enke}, {Eriksson}, {Esquej}, {Eynard Bontemps}, {Fabre},
  {Fabrizio}, {Faigler}, {Falc{\~a}o}, {Farr{\`a}s Casas}, {Federici},
  {Fedorets}, {Fernique}, {Figueras}, {Filippi}, {Findeisen}, {Fonti},
  {Fraile}, {Fraser}, {Fr{\'e}zouls}, {Gai}, {Galleti}, {Garabato},
  {Garc{\'\i}a-Sedano}, {Garofalo}, {Garralda}, {Gavel}, {Gavras}, {Gerssen},
  {Geyer}, {Giacobbe}, {Gilmore}, {Girona}, {Giuffrida}, {Glass}, {Gomes},
  {Granvik}, {Gueguen}, {Guerrier}, {Guiraud}, {Guti{\'e}rrez-S{\'a}nchez},
  {Haigron}, {Hatzidimitriou}, {Hauser}, {Haywood}, {Heiter}, {Helmi}, {Heu},
  {Hilger}, {Hobbs}, {Hofmann}, {Holland}, {Huckle}, {Hypki}, {Icardi},
  {Jan{\ss}en}, {Jevardat de Fombelle}, {Jonker}, {Juh{\'a}sz}, {Julbe},
  {Karampelas}, {Kewley}, {Klar}, {Kochoska}, {Kohley}, {Kolenberg},
  {Kontizas}, {Kontizas}, {Koposov}, {Kordopatis}, {Kostrzewa-Rutkowska},
  {Koubsky}, {Lambert}, {Lanza}, {Lasne}, {Lavigne}, {Le Fustec}, {Le
  Poncin-Lafitte}, {Lebreton}, {Leccia}, {Leclerc}, {Lecoeur-Taibi},
  {Lenhardt}, {Leroux}, {Liao}, {Licata}, {Lindstr{\o}m}, {Lister}, {Livanou},
  {Lobel}, {L{\'o}pez}, {Managau}, {Mann}, {Mantelet}, {Marchal}, {Marchant},
  {Marconi}, {Marinoni}, {Marschalk{\'o}}, {Marshall}, {Martino}, {Marton},
  {Mary}, {Massari}, {Matijevi{\v{c}}}, {Mazeh}, {McMillan}, {Messina},
  {Michalik}, {Millar}, {Molina}, {Molinaro}, {Moln{\'a}r}, {Montegriffo},
  {Mor}, {Morbidelli}, {Morel}, {Morris}, {Mulone}, {Muraveva}, {Musella},
  {Nelemans}, {Nicastro}, {Noval}, {O'Mullane}, {Ord{\'e}novic},
  {Ord{\'o}{\~n}ez-Blanco}, {Osborne}, {Pagani}, {Pagano}, {Pailler},
  {Palacin}, {Palaversa}, {Panahi}, {Pawlak}, {Piersimoni}, {Pineau}, {Plachy},
  {Plum}, {Poggio}, {Poujoulet}, {Pr{\v{s}}a}, {Pulone}, {Racero}, {Ragaini},
  {Rambaux}, {Ramos-Lerate}, {Regibo}, {Reyl{\'e}}, {Riclet}, {Ripepi}, {Riva},
  {Rivard}, {Rixon}, {Roegiers}, {Roelens}, {Romero-G{\'o}mez}, {Rowell},
  {Royer}, {Ruiz-Dern}, {Sadowski}, {Sagrist{\`a} Sell{\'e}s}, {Sahlmann},
  {Salgado}, {Salguero}, {Sanna}, {Santana-Ros}, {Sarasso}, {Savietto},
  {Schultheis}, {Sciacca}, {Segol}, {Segovia}, {S{\'e}gransan}, {Shih},
  {Siltala}, {Silva}, {Smart}, {Smith}, {Solano}, {Solitro}, {Sordo}, {Soria
  Nieto}, {Souchay}, {Spagna}, {Spoto}, {Stampa}, {Steele},
  {Steidelm{\"u}ller}, {Stephenson}, {Stoev}, {Suess}, {Surdej}, {Szabados},
  {Szegedi-Elek}, {Tapiador}, {Taris}, {Tauran}, {Taylor}, {Teixeira},
  {Terrett}, {Teyssandier}, {Thuillot}, {Titarenko}, {Torra Clotet}, {Turon},
  {Ulla}, {Utrilla}, {Uzzi}, {Vaillant}, {Valentini}, {Valette}, {van Elteren},
  {Van Hemelryck}, {van Leeuwen}, {Vaschetto}, {Vecchiato}, {Veljanoski},
  {Viala}, {Vicente}, {Vogt}, {von Essen}, {Voss}, {Votruba}, {Voutsinas},
  {Walmsley}, {Weiler}, {Wertz}, {Wevers}, {Wyrzykowski}, {Yoldas},
  {{\v{Z}}erjal}, {Ziaeepour}, {Zorec}, {Zschocke}, {Zucker}, {Zurbach}, \&
  {Zwitter}}]{gaia18}
{Gaia Collaboration}, {Brown}, A.~G.~A., {Vallenari}, A., {et~al.} 2018, \aap,
  616, A1

\bibitem[{{Gaia Collaboration} {et~al.}(2016){Gaia Collaboration}, {Prusti},
  {de Bruijne}, {Brown}, {Vallenari}, {Babusiaux}, {Bailer-Jones}, {Bastian},
  {Biermann}, {Evans}, {Eyer}, {Jansen}, {Jordi}, {Klioner}, {Lammers},
  {Lindegren}, {Luri}, {Mignard}, {Milligan}, {Panem}, {Poinsignon},
  {Pourbaix}, {Randich}, {Sarri}, {Sartoretti}, {Siddiqui}, {Soubiran},
  {Valette}, {van Leeuwen}, {Walton}, {Aerts}, {Arenou}, {Cropper}, {Drimmel},
  {H{\o}g}, {Katz}, {Lattanzi}, {O'Mullane}, {Grebel}, {Holland}, {Huc},
  {Passot}, {Bramante}, {Cacciari}, {Casta{\~n}eda}, {Chaoul}, {Cheek}, {De
  Angeli}, {Fabricius}, {Guerra}, {Hern{\'a}ndez}, {Jean-Antoine-Piccolo},
  {Masana}, {Messineo}, {Mowlavi}, {Nienartowicz}, {Ord{\'o}{\~n}ez-Blanco},
  {Panuzzo}, {Portell}, {Richards}, {Riello}, {Seabroke}, {Tanga},
  {Th{\'e}venin}, {Torra}, {Els}, {Gracia-Abril}, {Comoretto},
  {Garcia-Reinaldos}, {Lock}, {Mercier}, {Altmann}, {Andrae}, {Astraatmadja},
  {Bellas-Velidis}, {Benson}, {Berthier}, {Blomme}, {Busso}, {Carry},
  {Cellino}, {Clementini}, {Cowell}, {Creevey}, {Cuypers}, {Davidson}, {De
  Ridder}, {de Torres}, {Delchambre}, {Dell'Oro}, {Ducourant}, {Fr{\'e}mat},
  {Garc{\'\i}a-Torres}, {Gosset}, {Halbwachs}, {Hambly}, {Harrison}, {Hauser},
  {Hestroffer}, {Hodgkin}, {Huckle}, {Hutton}, {Jasniewicz}, {Jordan},
  {Kontizas}, {Korn}, {Lanzafame}, {Manteiga}, {Moitinho}, {Muinonen},
  {Osinde}, {Pancino}, {Pauwels}, {Petit}, {Recio-Blanco}, {Robin}, {Sarro},
  {Siopis}, {Smith}, {Smith}, {Sozzetti}, {Thuillot}, {van Reeven}, {Viala},
  {Abbas}, {Abreu Aramburu}, {Accart}, {Aguado}, {Allan}, {Allasia},
  {Altavilla}, {{\'A}lvarez}, {Alves}, {Anderson}, {Andrei}, {Anglada Varela},
  {Antiche}, {Antoja}, {Ant{\'o}n}, {Arcay}, {Atzei}, {Ayache}, {Bach},
  {Baker}, {Balaguer-N{\'u}{\~n}ez}, {Barache}, {Barata}, {Barbier}, {Barblan},
  {Baroni}, {Barrado y Navascu{\'e}s}, {Barros}, {Barstow}, {Becciani},
  {Bellazzini}, {Bellei}, {Bello Garc{\'\i}a}, {Belokurov}, {Bendjoya},
  {Berihuete}, {Bianchi}, {Bienaym{\'e}}, {Billebaud}, {Blagorodnova},
  {Blanco-Cuaresma}, {Boch}, {Bombrun}, {Borrachero}, {Bouquillon}, {Bourda},
  {Bouy}, {Bragaglia}, {Breddels}, {Brouillet}, {Br{\"u}semeister},
  {Bucciarelli}, {Budnik}, {Burgess}, {Burgon}, {Burlacu}, {Busonero}, {Buzzi},
  {Caffau}, {Cambras}, {Campbell}, {Cancelliere}, {Cantat-Gaudin}, {Carlucci},
  {Carrasco}, {Castellani}, {Charlot}, {Charnas}, {Charvet}, {Chassat},
  {Chiavassa}, {Clotet}, {Cocozza}, {Collins}, {Collins}, {Costigan}, {Crifo},
  {Cross}, {Crosta}, {Crowley}, {Dafonte}, {Damerdji}, {Dapergolas}, {David},
  {David}, {De Cat}, {de Felice}, {de Laverny}, {De Luise}, {De March}, {de
  Martino}, {de Souza}, {Debosscher}, {del Pozo}, {Delbo}, {Delgado},
  {Delgado}, {di Marco}, {Di Matteo}, {Diakite}, {Distefano}, {Dolding}, {Dos
  Anjos}, {Drazinos}, {Dur{\'a}n}, {Dzigan}, {Ecale}, {Edvardsson}, {Enke},
  {Erdmann}, {Escolar}, {Espina}, {Evans}, {Eynard Bontemps}, {Fabre},
  {Fabrizio}, {Faigler}, {Falc{\~a}o}, {Farr{\`a}s Casas}, {Faye}, {Federici},
  {Fedorets}, {Fern{\'a}ndez-Hern{\'a}ndez}, {Fernique}, {Fienga}, {Figueras},
  {Filippi}, {Findeisen}, {Fonti}, {Fouesneau}, {Fraile}, {Fraser}, {Fuchs},
  {Furnell}, {Gai}, {Galleti}, {Galluccio}, {Garabato}, {Garc{\'\i}a-Sedano},
  {Gar{\'e}}, {Garofalo}, {Garralda}, {Gavras}, {Gerssen}, {Geyer}, {Gilmore},
  {Girona}, {Giuffrida}, {Gomes}, {Gonz{\'a}lez-Marcos},
  {Gonz{\'a}lez-N{\'u}{\~n}ez}, {Gonz{\'a}lez-Vidal}, {Granvik}, {Guerrier},
  {Guillout}, {Guiraud}, {G{\'u}rpide}, {Guti{\'e}rrez-S{\'a}nchez}, {Guy},
  {Haigron}, {Hatzidimitriou}, {Haywood}, {Heiter}, {Helmi}, {Hobbs},
  {Hofmann}, {Holl}, {Holland}, {Hunt}, {Hypki}, {Icardi}, {Irwin}, {Jevardat
  de Fombelle}, {Jofr{\'e}}, {Jonker}, {Jorissen}, {Julbe}, {Karampelas},
  {Kochoska}, {Kohley}, {Kolenberg}, {Kontizas}, {Koposov}, {Kordopatis},
  {Koubsky}, {Kowalczyk}, {Krone-Martins}, {Kudryashova}, {Kull}, {Bachchan},
  {Lacoste-Seris}, {Lanza}, {Lavigne}, {Le Poncin-Lafitte}, {Lebreton},
  {Lebzelter}, {Leccia}, {Leclerc}, {Lecoeur-Taibi}, {Lemaitre}, {Lenhardt},
  {Leroux}, {Liao}, {Licata}, {Lindstr{\o}m}, {Lister}, {Livanou}, {Lobel},
  {L{\"o}ffler}, {L{\'o}pez}, {Lopez-Lozano}, {Lorenz}, {Loureiro},
  {MacDonald}, {Magalh{\~a}es Fernandes}, {Managau}, {Mann}, {Mantelet},
  {Marchal}, {Marchant}, {Marconi}, {Marie}, {Marinoni}, {Marrese},
  {Marschalk{\'o}}, {Marshall}, {Mart{\'\i}n-Fleitas}, {Martino}, {Mary},
  {Matijevi{\v{c}}}, {Mazeh}, {McMillan}, {Messina}, {Mestre}, {Michalik},
  {Millar}, {Miranda}, {Molina}, {Molinaro}, {Molinaro}, {Moln{\'a}r},
  {Moniez}, {Montegriffo}, {Monteiro}, {Mor}, {Mora}, {Morbidelli}, {Morel},
  {Morgenthaler}, {Morley}, {Morris}, {Mulone}, {Muraveva}, {Musella},
  {Narbonne}, {Nelemans}, {Nicastro}, {Noval}, {Ord{\'e}novic},
  {Ordieres-Mer{\'e}}, {Osborne}, {Pagani}, {Pagano}, {Pailler}, {Palacin},
  {Palaversa}, {Parsons}, {Paulsen}, {Pecoraro}, {Pedrosa}, {Pentik{\"a}inen},
  {Pereira}, {Pichon}, {Piersimoni}, {Pineau}, {Plachy}, {Plum}, {Poujoulet},
  {Pr{\v{s}}a}, {Pulone}, {Ragaini}, {Rago}, {Rambaux}, {Ramos-Lerate},
  {Ranalli}, {Rauw}, {Read}, {Regibo}, {Renk}, {Reyl{\'e}}, {Ribeiro},
  {Rimoldini}, {Ripepi}, {Riva}, {Rixon}, {Roelens}, {Romero-G{\'o}mez},
  {Rowell}, {Royer}, {Rudolph}, {Ruiz-Dern}, {Sadowski}, {Sagrist{\`a}
  Sell{\'e}s}, {Sahlmann}, {Salgado}, {Salguero}, {Sarasso}, {Savietto},
  {Schnorhk}, {Schultheis}, {Sciacca}, {Segol}, {Segovia}, {Segransan},
  {Serpell}, {Shih}, {Smareglia}, {Smart}, {Smith}, {Solano}, {Solitro},
  {Sordo}, {Soria Nieto}, {Souchay}, {Spagna}, {Spoto}, {Stampa}, {Steele},
  {Steidelm{\"u}ller}, {Stephenson}, {Stoev}, {Suess}, {S{\"u}veges}, {Surdej},
  {Szabados}, {Szegedi-Elek}, {Tapiador}, {Taris}, {Tauran}, {Taylor},
  {Teixeira}, {Terrett}, {Tingley}, {Trager}, {Turon}, {Ulla}, {Utrilla},
  {Valentini}, {van Elteren}, {Van Hemelryck}, {van Leeuwen}, {Varadi},
  {Vecchiato}, {Veljanoski}, {Via}, {Vicente}, {Vogt}, {Voss}, {Votruba},
  {Voutsinas}, {Walmsley}, {Weiler}, {Weingrill}, {Werner}, {Wevers},
  {Whitehead}, {Wyrzykowski}, {Yoldas}, {{\v{Z}}erjal}, {Zucker}, {Zurbach},
  {Zwitter}, {Alecu}, {Allen}, {Allende Prieto}, {Amorim},
  {Anglada-Escud{\'e}}, {Arsenijevic}, {Azaz}, {Balm}, {Beck}, {Bernstein},
  {Bigot}, {Bijaoui}, {Blasco}, {Bonfigli}, {Bono}, {Boudreault}, {Bressan},
  {Brown}, {Brunet}, {Bunclark}, {Buonanno}, {Butkevich}, {Carret}, {Carrion},
  {Chemin}, {Ch{\'e}reau}, {Corcione}, {Darmigny}, {de Boer}, {de Teodoro}, {de
  Zeeuw}, {Delle Luche}, {Domingues}, {Dubath}, {Fodor}, {Fr{\'e}zouls},
  {Fries}, {Fustes}, {Fyfe}, {Gallardo}, {Gallegos}, {Gardiol}, {Gebran},
  {Gomboc}, {G{\'o}mez}, {Grux}, {Gueguen}, {Heyrovsky}, {Hoar}, {Iannicola},
  {Isasi Parache}, {Janotto}, {Joliet}, {Jonckheere}, {Keil}, {Kim},
  {Klagyivik}, {Klar}, {Knude}, {Kochukhov}, {Kolka}, {Kos}, {Kutka}, {Lainey},
  {LeBouquin}, {Liu}, {Loreggia}, {Makarov}, {Marseille}, {Martayan},
  {Martinez-Rubi}, {Massart}, {Meynadier}, {Mignot}, {Munari}, {Nguyen},
  {Nordlander}, {Ocvirk}, {O'Flaherty}, {Olias Sanz}, {Ortiz}, {Osorio},
  {Oszkiewicz}, {Ouzounis}, {Palmer}, {Park}, {Pasquato}, {Peltzer}, {Peralta},
  {P{\'e}turaud}, {Pieniluoma}, {Pigozzi}, {Poels}, {Prat}, {Prod'homme},
  {Raison}, {Rebordao}, {Risquez}, {Rocca-Volmerange}, {Rosen}, {Ruiz-Fuertes},
  {Russo}, {Sembay}, {Serraller Vizcaino}, {Short}, {Siebert}, {Silva},
  {Sinachopoulos}, {Slezak}, {Soffel}, {Sosnowska}, {Strai{\v{z}}ys}, {ter
  Linden}, {Terrell}, {Theil}, {Tiede}, {Troisi}, {Tsalmantza}, {Tur},
  {Vaccari}, {Vachier}, {Valles}, {Van Hamme}, {Veltz}, {Virtanen}, {Wallut},
  {Wichmann}, {Wilkinson}, {Ziaeepour}, \& {Zschocke}}]{gaia16}
{Gaia Collaboration}, {Prusti}, T., {de Bruijne}, J.~H.~J., {et~al.} 2016,
  \aap, 595, A1

\bibitem[{Ghosh \& Lamb(1979)}]{gho79}
Ghosh, P. \& Lamb, F. 1979, The Astrophysical Journal, 232, 259

\bibitem[{Gressel {et~al.}(2013)Gressel, Nelson, Turner, \& Ziegler}]{gre13}
Gressel, O., Nelson, R.~P., Turner, N.~J., \& Ziegler, U. 2013, The
  Astrophysical Journal, 779, 59

\bibitem[{Haffert {et~al.}(2019)Haffert, Bohn, de~Boer, Snellen, Brinchmann,
  Girard, Keller, \& Bacon}]{haf19}
Haffert, S., Bohn, A., de~Boer, J., {et~al.} 2019, Nature Astronomy, 3, 749

\bibitem[{Hashimoto {et~al.}(2020)Hashimoto, Aoyama, Konishi, Uyama, Takasao,
  Ikoma, \& Tanigawa}]{has20}
Hashimoto, J., Aoyama, Y., Konishi, M., {et~al.} 2020, The Astronomical
  Journal, 159, 222

\bibitem[{Hashimoto {et~al.}(2015)Hashimoto, Tsukagoshi, Brown, Dong, Muto,
  Zhu, Wisniewski, Ohashi, Kusakabe, Abe, {et~al.}}]{has15}
Hashimoto, J., Tsukagoshi, T., Brown, J.~M., {et~al.} 2015, The Astrophysical
  Journal, 799, 43

\bibitem[{Homma {et~al.}(2020)Homma, Ohtsuki, Maeda, Suetsugu, Machida, \&
  Tanigawa}]{hom20}
Homma, T., Ohtsuki, K., Maeda, N., {et~al.} 2020, The Astrophysical Journal

\bibitem[{Hueso \& Guillot(2005)}]{hue05}
Hueso, R. \& Guillot, T. 2005, Astronomy \& Astrophysics, 442, 703

\bibitem[{Hyodo {et~al.}(2021)Hyodo, Guillot, Ida, Okuzumi, \& Youdin}]{hyo21a}
Hyodo, R., Guillot, T., Ida, S., Okuzumi, S., \& Youdin, A.~N. 2021, Astronomy
  \& Astrophysics, 646, A14

\bibitem[{Isella {et~al.}(2019)Isella, Benisty, Teague, Bae, Keppler, Facchini,
  \& P{\'e}rez}]{ise19}
Isella, A., Benisty, M., Teague, R., {et~al.} 2019, The Astrophysical Journal
  Letters, 879, L25

\bibitem[{Kanagawa {et~al.}(2018)Kanagawa, Muto, Okuzumi, Tanigawa, Taki, \&
  Shibaike}]{kan18}
Kanagawa, K.~D., Muto, T., Okuzumi, S., {et~al.} 2018, The Astrophysical
  Journal, 868, 48

\bibitem[{Kanagawa {et~al.}(2016)Kanagawa, Muto, Tanaka, Tanigawa, Takeuchi,
  Tsukagoshi, \& Momose}]{kan16}
Kanagawa, K.~D., Muto, T., Tanaka, H., {et~al.} 2016, Publications of the
  Astronomical Society of Japan, 68, 43

\bibitem[{Karlin {et~al.}(2023)Karlin, Pani{\'c}, \& van Loo}]{kar23}
Karlin, S.~M., Pani{\'c}, O., \& van Loo, S. 2023, Monthly Notices of the Royal
  Astronomical Society, 520, 1258

\bibitem[{Kataoka {et~al.}(2014)Kataoka, Okuzumi, Tanaka, \& Nomura}]{kat14}
Kataoka, A., Okuzumi, S., Tanaka, H., \& Nomura, H. 2014, Astronomy \&
  Astrophysics, 568, A42

\bibitem[{Keppler {et~al.}(2018)Keppler, Benisty, M{\"u}ller, Henning,
  Van~Boekel, Cantalloube, Ginski, Van~Holstein, Maire, Pohl, {et~al.}}]{kep18}
Keppler, M., Benisty, M., M{\"u}ller, A., {et~al.} 2018, Astronomy \&
  Astrophysics, 617, A44

\bibitem[{Keppler {et~al.}(2019)Keppler, Teague, Bae, Benisty, Henning,
  Van~Boekel, Chapillon, Pinilla, Williams, Bertrang, {et~al.}}]{kep19}
Keppler, M., Teague, R., Bae, J., {et~al.} 2019, Astronomy \& Astrophysics,
  625, A118

\bibitem[{Kratter \& Lodato(2016)}]{kra16}
Kratter, K. \& Lodato, G. 2016, Annual Review of Astronomy and Astrophysics,
  54, 271

\bibitem[{Law {et~al.}(2024)Law, Benisty, Facchini, Teague, Bae, Isella, Kamp,
  {\"O}berg, Portilla-Revelo, \& Rampinelli}]{law24}
Law, C.~J., Benisty, M., Facchini, S., {et~al.} 2024, arXiv preprint
  arXiv:2401.03018

\bibitem[{Lissauer \& Kary(1991)}]{lis91}
Lissauer, J.~J. \& Kary, D.~M. 1991, Icarus, 94, 126

\bibitem[{Lubow \& D'Angelo(2006)}]{lub06}
Lubow, S.~H. \& D'Angelo, G. 2006, The Astrophysical Journal, 641, 526

\bibitem[{Lubow {et~al.}(1999)Lubow, Seibert, \& Artymowicz}]{lub99}
Lubow, S.~H., Seibert, M., \& Artymowicz, P. 1999, The Astrophysical Journal,
  526, 1001

\bibitem[{Marleau {et~al.}(2022)Marleau, Aoyama, Kuiper, Follette, Turner,
  Cugno, Manara, Haffert, Kitzmann, Ringqvist, {et~al.}}]{mar22}
Marleau, G.-D., Aoyama, Y., Kuiper, R., {et~al.} 2022, Astronomy \&
  Astrophysics, 657, A38

\bibitem[{Marleau {et~al.}(2023)Marleau, Kuiper, B{\'e}thune, \&
  Mordasini}]{mar23}
Marleau, G.-D., Kuiper, R., B{\'e}thune, W., \& Mordasini, C. 2023, The
  Astrophysical Journal, 952, 89

\bibitem[{Marleau {et~al.}(2019)Marleau, Mordasini, \& Kuiper}]{mar19}
Marleau, G.-D., Mordasini, C., \& Kuiper, R. 2019, The Astrophysical Journal,
  881, 144

\bibitem[{Michikoshi \& Inutsuka(2006)}]{mic06}
Michikoshi, S. \& Inutsuka, S.-i. 2006, The Astrophysical Journal, 641, 1131

\bibitem[{M{\"u}ller {et~al.}(2018)M{\"u}ller, Keppler, Henning, Samland,
  Chauvin, Beust, Maire, Molaverdikhani, van Boekel, Benisty, {et~al.}}]{mul18}
M{\"u}ller, A., Keppler, M., Henning, T., {et~al.} 2018, Astronomy \&
  Astrophysics, 617, L2

\bibitem[{Muzerolle {et~al.}(2001)Muzerolle, Calvet, \& Hartmann}]{muz01}
Muzerolle, J., Calvet, N., \& Hartmann, L. 2001, The Astrophysical Journal,
  550, 944

\bibitem[{Nakamoto \& Nakagawa(1994)}]{nak94}
Nakamoto, T. \& Nakagawa, Y. 1994, The Astrophysical Journal, 421, 640

\bibitem[{Natta {et~al.}(2004)Natta, Testi, Muzerolle, Randich, Comer{\'o}n, \&
  Persi}]{nat04}
Natta, A., Testi, L., Muzerolle, J., {et~al.} 2004, Astronomy \& Astrophysics,
  424, 603

\bibitem[{Okuzumi {et~al.}(2016)Okuzumi, Momose, iti Sirono, Kobayashi, \&
  Tanaka}]{oku16}
Okuzumi, S., Momose, M., iti Sirono, S., Kobayashi, H., \& Tanaka, H. 2016, The
  Astrophysical Journal, 821, 82

\bibitem[{Okuzumi {et~al.}(2012)Okuzumi, Tanaka, Kobayashi, \& Wada}]{oku12}
Okuzumi, S., Tanaka, H., Kobayashi, H., \& Wada, K. 2012, The Astrophysical
  Journal, 752, 106

\bibitem[{Ormel \& Cuzzi(2007)}]{orm07}
Ormel, C. \& Cuzzi, J. 2007, Astronomy \& Astrophysics, 466, 413

\bibitem[{Perets \& Murray-Clay(2011)}]{per11}
Perets, H.~B. \& Murray-Clay, R.~A. 2011, The Astrophysical Journal, 733, 56

\bibitem[{Pollack {et~al.}(1994)Pollack, Hollenbach, Beckwith, Simonelli,
  Roush, \& Fong}]{pol94}
Pollack, J.~B., Hollenbach, D., Beckwith, S., {et~al.} 1994, The Astrophysical
  Journal, 421, 615

\bibitem[{Portilla-Revelo {et~al.}(2023)Portilla-Revelo, Kamp, Facchini,
  Van~Dishoeck, Law, Rab, Bae, Benisty, {\"O}berg, \& Teague}]{por23}
Portilla-Revelo, B., Kamp, I., Facchini, S., {et~al.} 2023, Astronomy \&
  Astrophysics, 677, A76

\bibitem[{Rigliaco {et~al.}(2012)Rigliaco, Natta, Testi, Randich, Alcala,
  Covino, \& Stelzer}]{rig12}
Rigliaco, E., Natta, A., Testi, L., {et~al.} 2012, Astronomy \& Astrophysics,
  548, A56

\bibitem[{Ronnet {et~al.}(2017)Ronnet, Mousis, \& Vernazza}]{ron17}
Ronnet, T., Mousis, O., \& Vernazza, P. 2017, The Astrophysical Journal, 845,
  92

\bibitem[{Sato {et~al.}(2016)Sato, Okuzumi, \& Ida}]{sat16}
Sato, T., Okuzumi, S., \& Ida, S. 2016, Astronomy \& Astrophysics, 589, A15

\bibitem[{Schulik {et~al.}(2020)Schulik, Johansen, Bitsch, Lega, \&
  Lambrechts}]{sch20}
Schulik, M., Johansen, A., Bitsch, B., Lega, E., \& Lambrechts, M. 2020,
  Astronomy \& Astrophysics, 642, A187

\bibitem[{{Shakura} \& {Sunyaev}(1973)}]{sha73}
{Shakura}, N.~I. \& {Sunyaev}, R.~A. 1973, Astronomy \& Astrophysics, 24, 337

\bibitem[{Shibaike \& Mori(2023)}]{shi23}
Shibaike, Y. \& Mori, S. 2023, Monthly Notices of the Royal Astronomical
  Society, 518, 5444

\bibitem[{Shibaike {et~al.}(2017)Shibaike, Okuzumi, Sasaki, \& Ida}]{shi17}
Shibaike, Y., Okuzumi, S., Sasaki, T., \& Ida, S. 2017, The Astrophysical
  Journal, 846, 10pp

\bibitem[{Shibaike {et~al.}(2019)Shibaike, Ormel, Ida, Okuzumi, \&
  Sasaki}]{shi19}
Shibaike, Y., Ormel, C.~W., Ida, S., Okuzumi, S., \& Sasaki, T. 2019, The
  Astrophysical Journal, 885, 79

\bibitem[{Stolker {et~al.}(2020)Stolker, Marleau, Cugno, Molli{\`e}re, Quanz,
  Todorov, \& K{\"u}hn}]{sto20}
Stolker, T., Marleau, G.~D., Cugno, G., {et~al.} 2020, Astronomy \&
  Astrophysics, 644, A13

\bibitem[{Szul{\'a}gyi {et~al.}(2022)Szul{\'a}gyi, Binkert, \&
  Surville}]{szu22}
Szul{\'a}gyi, J., Binkert, F., \& Surville, C. 2022, The Astrophysical Journal,
  924, 1

\bibitem[{Takasao {et~al.}(2021)Takasao, Aoyama, \& Ikoma}]{tak21}
Takasao, S., Aoyama, Y., \& Ikoma, M. 2021, The Astrophysical Journal, 921, 10

\bibitem[{Takasao {et~al.}(2022)Takasao, Tomida, Iwasaki, \& Suzuki}]{tak22}
Takasao, S., Tomida, K., Iwasaki, K., \& Suzuki, T.~K. 2022, The Astrophysical
  Journal, 941, 73

\bibitem[{Tanigawa {et~al.}(2012)Tanigawa, Ohtsuki, \& Machida}]{tan12}
Tanigawa, T., Ohtsuki, K., \& Machida, M.~N. 2012, The Astrophysical Journal,
  747, 47

\bibitem[{Thanathibodee {et~al.}(2019)Thanathibodee, Calvet, Bae, Muzerolle, \&
  Hern{\'a}ndez}]{tha19}
Thanathibodee, T., Calvet, N., Bae, J., Muzerolle, J., \& Hern{\'a}ndez, R.~F.
  2019, The Astrophysical Journal, 885, 94

\bibitem[{Thanathibodee {et~al.}(2020)Thanathibodee, Molina, Calvet, Serna,
  Bae, Reynolds, Hern{\'a}ndez, Muzerolle, \& Hern{\'a}ndez}]{tha20}
Thanathibodee, T., Molina, B., Calvet, N., {et~al.} 2020, The Astrophysical
  Journal, 892, 81

\bibitem[{Toomre(1964)}]{too64}
Toomre, A. 1964, Astrophysical Journal, vol. 139, p. 1217-1238 (1964)., 139,
  1217

\bibitem[{Turner {et~al.}(2014)Turner, Lee, \& Sano}]{tur14}
Turner, N.~J., Lee, M.~H., \& Sano, T. 2014, The Astrophysical Journal, 783, 14

\bibitem[{{Wada} {et~al.}(2013){Wada}, {Tanaka}, {Okuzumi}, {Kobayashi},
  {Suyama}, {Kimura}, \& {Yamamoto}}]{wad13}
{Wada}, K., {Tanaka}, H., {Okuzumi}, S., {et~al.} 2013, \aap, 559, A62

\bibitem[{Wagner {et~al.}(2018)Wagner, Follette, Close, Apai, Gibbs, Keppler,
  M{\"u}ller, Henning, Kasper, Wu, {et~al.}}]{wag18}
Wagner, K., Follette, K.~B., Close, L.~M., {et~al.} 2018, The Astrophysical
  Journal Letters, 863, L8

\bibitem[{Wang {et~al.}(2021)Wang, Vigan, Lacour, Nowak, Stolker, De~Rosa,
  Ginzburg, Gao, Abuter, Amorim, {et~al.}}]{wan21}
Wang, J., Vigan, A., Lacour, S., {et~al.} 2021, The Astronomical Journal, 161,
  148

\bibitem[{Ward \& Canup(2010)}]{war10}
Ward, W.~R. \& Canup, R.~M. 2010, The Astronomical Journal, 140, 1168

\bibitem[{Weidenschilling(1977)}]{wei77}
Weidenschilling, S. 1977, Icarus, 44, 172

\bibitem[{Whipple(1972)}]{whi72}
Whipple, F. 1972, From plasma to planet, ed. A. Elvius (London: Wiley)

\bibitem[{Youdin \& Lithwick(2007)}]{you07}
Youdin, A.~N. \& Lithwick, Y. 2007, Icarus, 192, 588

\bibitem[{Zhou {et~al.}(2021)Zhou, Bowler, Wagner, Schneider, Apai, Kraus,
  Close, Herczeg, \& Fang}]{zho21}
Zhou, Y., Bowler, B.~P., Wagner, K.~R., {et~al.} 2021, The Astronomical
  Journal, 161, 244

\bibitem[{Zhu(2015)}]{zhu15a}
Zhu, Z. 2015, The Astrophysical Journal, 799, 16

\bibitem[{Zhu {et~al.}(2016)Zhu, Ju, \& Stone}]{zhu16}
Zhu, Z., Ju, W., \& Stone, J.~M. 2016, The Astrophysical Journal, 832, 193

\bibitem[{Zhu {et~al.}(2012)Zhu, Nelson, Dong, Espaillat, \& Hartmann}]{zhu12}
Zhu, Z., Nelson, R.~P., Dong, R., Espaillat, C., \& Hartmann, L. 2012, The
  Astrophysical Journal, 755, 6

\end{thebibliography}


\begin{appendix} 
\section{Approximations of the dust-to-gas surface density ratio for the opacity calculation} \label{opacity}
\begin{figure}[t]
\centering
\includegraphics[width=0.9\linewidth]{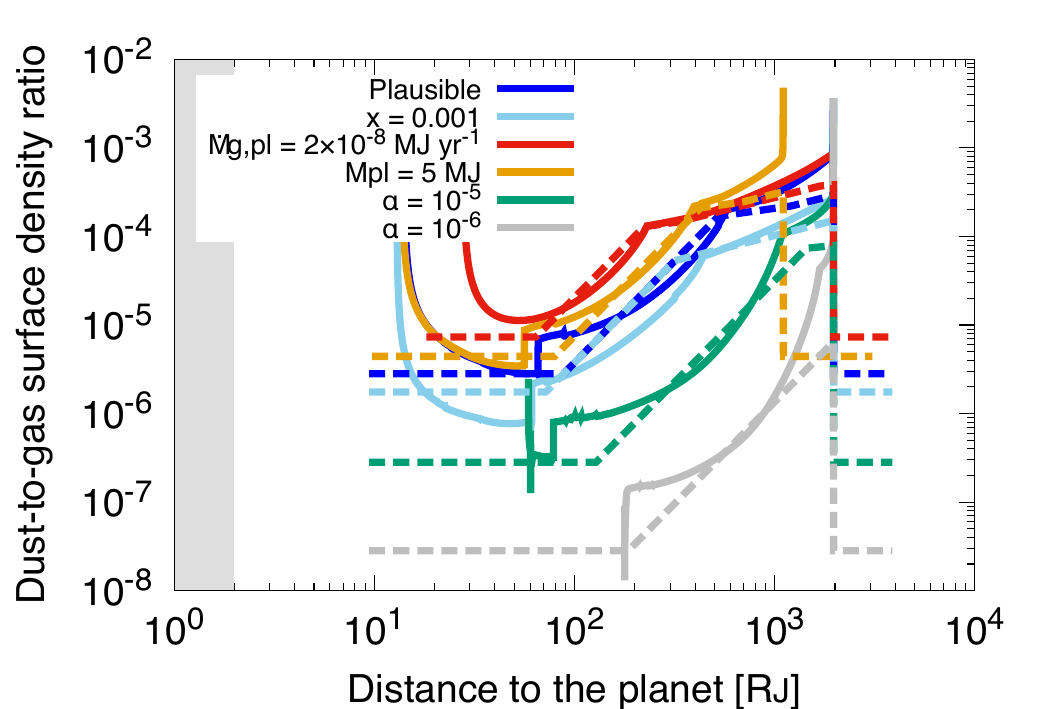}
\caption{Radial distribution of $Z_{\rm\Sigma}$ (solid curves) and $Z_{\rm\Sigma,est}$ (dashed lines) with the various parameter sets investigated in Section \ref{detailed}.}
\label{fig:opacity}
\end{figure}

In this work, we calculate the radial distribution of the surface density and size of dust, but the opacity used for the calculations of the gas disc structures is not calculated simultaneously. We estimate the radial distribution of dust-to-gas surface density ratio, $Z_{\rm\Sigma,est}$, by the combination of three power-low approximations directly obtained from the input parameters. First, the estimate can be divided by inside or outside the snowline, we assume
\begin{equation}
Z_{\rm\Sigma,est}=\max(Z_{\rm\Sigma,est,ice}, Z_{\rm\Sigma,est,rock}).
\label{ZSigmaest}
\end{equation}
Then, outside the snowline, the particles are in the Epstein regime when they are supplied to CPDs and move to the Stokes regime, we assume
\begin{equation}
Z_{\rm\Sigma,est,ice}=\min(Z_{\rm\Sigma,est,ice,Ep}, Z_{\rm\Sigma,est,ice,St}).
\label{ZSigmaice}
\end{equation}
Inside the snowline, the particles are always in the Stokes regime,
\begin{equation}
Z_{\rm\Sigma,est,rock}=Z_{\rm\Sigma,est,rock,St}.
\label{ZSigmarock}
\end{equation}
Considering that whether the collision velocity is driven by diffusion (turbulence) or drift depends on the strength of turbulence, we assume
\begin{equation}
Z_{\rm\Sigma,est,ice,Ep}=\min(Z_{\rm\Sigma,est,ice,Ep,vr}, Z_{\rm\Sigma,est,ice,Ep,vt}),
\label{ZSigmaiceEp}
\end{equation}
\begin{equation}
Z_{\rm\Sigma,est,ice,St}=\min(Z_{\rm\Sigma,est,ice,St,vr}, Z_{\rm\Sigma,est,ice,St,vt}),
\label{ZSigmaiceSt}
\end{equation}
and
\begin{equation}
Z_{\rm\Sigma,est,rock,St}=\min(Z_{\rm\Sigma,est,rock,St,vr}, Z_{\rm\Sigma,est,rock,St,vt}).
\label{ZSigmarockSt}
\end{equation}
We calculate $Z_{\rm\Sigma,est,ice,Ep,vt}$ depending on which of the expressions of Eq. (\ref{vt}) determines the collision velocity,
\begin{equation}
Z_{\rm\Sigma,est,ice,Ep,vt}=\min(Z_{\rm\Sigma,est,ice,Ep,vt1}, Z_{\rm\Sigma,est,ice,Ep,vt2}),
\label{ZSigmaiceEpvt}
\end{equation}
where $Z_{\rm\Sigma,est,ice,Ep,vt1}$ and $Z_{\rm\Sigma,est,ice,Ep,vt2}$ correspond to the upper and middle expressions of Eq. (\ref{vt}), respectively. We use the middle expression for $Z_{\rm\Sigma,est,ice,St,vt}$ and $Z_{\rm\Sigma,est,rock,St,vt}$. The particles start to drift when the drift timescale becomes shorter than the growth timescale. Therefore, we assume that the drift and growth timescales of the drifting particles are equal, $r/|v_{\rm r}|=m_{\rm d}/|dm_{\rm d}/dt|$. We also consider a simple viscous accretion disc, then $\Sigma_{\rm g}=\dot{M}_{\rm g,tot}\Omega_{\rm K}/(3\pi\alpha c_{\rm s}^{2})$, where $\dot{M}_{\rm g,tot}$ is uniform. We also assume that $\dot{M}_{\rm d}=\dot{M}_{\rm d,tot}$ (uniform), $H_{\rm d}=H_{\rm g}\sqrt{\alpha/{\rm St}}$, and $|v_{\rm r}|=(H_{\rm g}/r)^{2}\gamma{\rm St}v_{\rm K}$, where $\gamma=\partial\ln{(\rho_{\rm g,mid}c_{\rm s}^{2})}/(\partial\ln{r})$ is a spacial constant. If the disc temperature is determined by viscous heating, it can be approximated to $T=(3\pi M_{\rm pl}\dot{M}_{\rm g,tot}/(8\sigma_{\rm SB})\times1/(4.8\tau_{\rm R}))^{1/4}$, because the disc is optically thin (see Fig. \ref{fig:evolution}). When the opacity is determined by the dust opacity, it can be assumed to $\kappa_{\rm R}\propto T^{2}Z_{\rm \Sigma}$ for outside the snowline, and $\kappa_{\rm R}\propto Z_{\rm \Sigma}$ for inside the snowline \citep{pol94}. Then, we obtain the dependence of $Z_{\rm \Sigma}$ on the planet and CPD properties,
\begin{equation}
\begin{split}
Z_{\rm\Sigma,est}&=Z_{\rm\Sigma,est,0}\left(\dfrac{\dot{M}_{\rm g,tot}}{2\times10^{-7}~M_{\rm J}~{\rm yr}^{-1}}\right)^{q_{\dot{M}_{\rm g,pl}}}\left(\dfrac{M_{\rm pl}}{10~M_{\rm pl}}\right)^{q_{M_{\rm pl}}} \\
&\left(\dfrac{\alpha}{10^{-4}}\right)^{q_{\alpha}}\times\left(\dfrac{x}0.01\right)^{q_{x}}\left(\dfrac{r}{100~R_{\rm J}}\right)^{q_{r}},
\end{split}
\label{ZSigmaest2}
\end{equation}
where the constants $Z_{\rm\Sigma,est,0}$, $q_{\dot{M}_{\rm g,tot}}$, $q_{M_{\rm pl}}$, $q_{\alpha}$, $q_{x}$, and $q_{r}$ are shown in Table \ref{tab:ZSigma}. Here, the $r$ dependence of $Z_{\rm\Sigma,est,ice,St}$ and $Z_{\rm\Sigma,est,rock,St}$ is assumed as $q_{r}=2.3$ and $q_{r}=0$, respectively, to approximate $Z_{\rm\Sigma}$ more correctly, which is not derived from the above discussion. The value of $Z_{\rm\Sigma,est,0}$ is not derived from it either but is just assumed. Figure \ref{fig:opacity} shows that this approximation, $Z_{\rm\Sigma,est}$, (dashed lines) reproduces $Z_{\rm\Sigma}$ (solid lines) well, especially outside the snowline. We note that the opacity is dominated by the gas opacity inside the snowline when the turbulence is weak (green and grey), so that the uncertainty of $Z_{\rm\Sigma}$ due to the growth of the dust inside the snowline should not be a significant problem.

\begin{table}[t]
\caption{Constants of $Z_{\rm\Sigma}$}
\label{tab:ZSigma}
\centering
\small
\begin{tabular}{ccccccc}
\hline
Description & $Z_{\rm\Sigma,est,0}$ & $q_{\dot{M}_{\rm g,pl}}$ & $q_{M_{\rm pl}}$ & $q_{\alpha}$ & $q_{x}$ & $q_{r}$ \\
\hline
\hline
$Z_{\rm\Sigma,est,ice,Ep,vr}$ & $2\times10^{-2}$ & $0$ & $-9/32$ & $11/16$ & $5/16$ & $7/32$ \\
$Z_{\rm\Sigma,est,ice,Ep,vt1}$ & $6\times10^{-3}$ & $-1/7$ & $-4/7$ & $3/7$ & $2/7$ & $4/7$ \\
$Z_{\rm\Sigma,est,ice,Ep,vt2}$ & $1\times10^{-2}$ & $0$ & $-9/22$ & $1/2$ & $5/11$ & $7/22$ \\
$Z_{\rm\Sigma,est,ice,St,vr}$ & $7\times10^{-4}$ & $-20/27$ & $-43/54$ & $37/27$ & $0$ & $2.3$ \\
$Z_{\rm\Sigma,est,ice,St,vt}$ & $1\times10^{-3}$ & $-20/41$ & $-59/82$ & $1$ & $10/41$ & $2.3$ \\
$Z_{\rm\Sigma,est,rock,St,vr}$ & $2\times10^{-4}$ & $-12/19$ & $-23/38$ & $25/19$ & $0$ & $0$ \\
$Z_{\rm\Sigma,est,rock,St,vt}$ & $2\times10^{-4}$ & $-12/29$ & $-31/58$ & $1$ & $6/29$ & $0$ \\
\hline
\end{tabular}
\end{table}

\section{Gravitational instability of the CPDs} \label{GI}
\begin{figure}[t]
\centering
\includegraphics[width=0.9\linewidth]{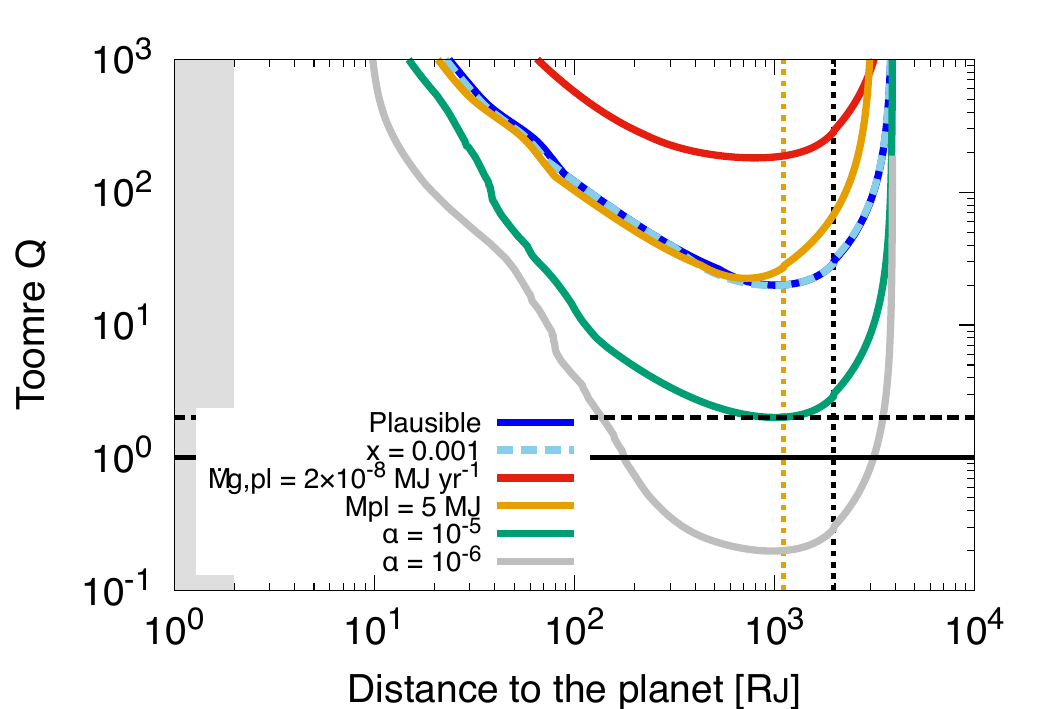}
\caption{Conditions for gravitational instability $Q_{\rm Toomre}$ with the various parameter sets in Section \ref{detailed}. The solid and dashed black horizontal lines are $Q_{\rm Toomre}=1$ and $2$, respectively.}
\label{fig:ToomreQ}
\end{figure}

When Toomre Q parameter, $Q_{\rm Toomre}$ (Eq. (\ref{ToomreQ})), is lower than unity, the gas disc is gravitationally unstable \citep[e.g.][]{too64}. In the case of the CPD of PDS~70~c with the plausible value of the planet mass and the gas accretion rate ($M_{\rm pl}=10~M_{\rm J}$ and $\dot{M}_{\rm g,pl}=2\times10^{-7}~M_{\rm J}~{\rm yr}^{-1}$), the condition for the instability is satisfied when the disc turbulence is very weak (about $\alpha<10^{-5}$) as shown in Fig. \ref{fig:parameters}. Figure \ref{fig:ToomreQ} represents the radial distribution of $Q_{\rm Toomre}$ and shows that the parameter is lower than unity at the outer region of the disc when $\alpha=10^{-6}$. Therefore, it should be difficult to keep the disc structure in such an extreme condition, especially at the outer region, but the prediction of the dust emission in this situation is beyond the scope of this paper. We also note that $Q_{\rm Toomre}$ is about $2$ around $r=1000~R_{\rm J}$ when $\alpha=10^{-5}$, which suggests non-axisymmetric features may develop at the region according to some previous research about protoplanetary discs \citep[e.g.][]{kra16}. However, it is unknown whether that is also the case in circumplanetary discs or not. It is difficult to describe non-axisymmetric features by our 1D disc model, but the necessary conditions obtained in this paper (i.e. the high gas accretion rate and weak turbulence) would be preserved even if such non-axisymmetric features are considered.

\section{Analytical solutions of optical depth} \label{analitical}
The optical depth for the wavelength of $\lambda$ in the model (Eq. (\ref{tau})) can be analytically solved. Substituting Eqs. (\ref{sfd}), (\ref{Sigma_dzero}), and (\ref{kappa_abs}) with Eq. (\ref{tau}),
\begin{equation}
\begin{split}
\tau_{\lambda}&=\dfrac{3\Sigma_{\rm d,0}}{4\rho_{\rm int,opa}}\int^{R_{\rm d}}_{a_{\rm min}}a^{2-q}Q_{\rm abs}~da, \\
&=
\begin{cases}
A_{\rm A} & \lambda_{0}>R_{\rm d}, \\
A_{\rm B} + A_{\rm C} & \lambda_{0}\leq R_{\rm d}.
\end{cases}
\end{split}
\label{tau-appendix}
\end{equation}
where $\lambda_{0}=2\pi a/\lambda$ and
\begin{equation}
\begin{split}
&A_{\rm A}=\int^{R_{\rm d}}_{a_{\rm min}}a^{2-q}Q_{\rm abs,1}~da, \\
&A_{\rm B}=\int^{\lambda_{0}}_{a_{\rm min}}a^{2-q}Q_{\rm abs,1}~da, \\
&A_{\rm C}=\int^{R_{\rm d}}_{\lambda_{0}}a^{2-q}\min{(Q_{\rm abs,2}, Q_{\rm abs,3})}~da.
\end{split}
\label{A_ABC}
\end{equation}
From Section \ref{emission}, we can express the coefficient $Q_{\rm abs}$ as $Q_{\rm abs,1}=(C_{1}/\lambda_{0})a$, $Q_{\rm abs,2}=(C_{2}/\lambda_{0})a$, and $Q_{\rm abs,3}=C_{3}$, where $C_{1}$, $C_{2}$, and $C_{3}$ are constants. Therefore, we get
\begin{equation}
A_{\rm A}=\dfrac{C_{1}}{\lambda_{\rm 0}(4-q)}(R_{\rm d}^{4-q}-a_{\rm min}^{4-q}),
\label{A_A}
\end{equation}
\begin{equation}
A_{\rm B}=\dfrac{C_{1}}{\lambda_{0}(4-q)}(\lambda_{0}^{4-q}-a_{\rm min}^{4-q}),
\label{A_B}
\end{equation}
and
\begin{equation}
A_{\rm C}=
\begin{cases}
\dfrac{C_{2}}{\lambda_{0}(4-q)}(R_{\rm d}^{4-q}-a_{\rm min}^{4-q}) & R_{\rm d}<a_{23}, \\
\dfrac{C_{3}}{\lambda_{0}(3-q)}(R_{\rm d}^{3-q}-a_{\rm min}^{3-q}) & \lambda_{0}>a_{23}, \\
\dfrac{C_{2}}{\lambda_{0}(4-q)}(a_{23}^{4-q}-\lambda_{0}^{4-q}) \\
+ \dfrac{C_{3}}{R_{\rm d}(3-q)}(R_{\rm d}^{3-q}-a_{23}^{3-q}) & {\rm otherwise},
\end{cases}
\label{A_C}
\end{equation}
where $a_{23}=(C_{3}/C_{2})\lambda_{0}$.

\end{appendix}

\end{document}